\documentclass{ws-ijmpa}
\usepackage{axodraw}
\newcommand{\ppbar}     {\mbox{$p\bar{p}$}}
\newcommand{\rargap}	{\mbox{ $\rightarrow$ }}
\begin{document}

\markboth{E.~Boos, L.~Dudko}
{The Single Top Quark Physics}

%
\catchline{}{}{}{}{}
%

\title{The Single Top Quark Physics}

\author{E.~Boos\footnote{boos@theory.sinp.msu.ru}, L.~Dudko\footnote{dudko@sinp.msu.ru}}

\address{Lomonosov Moscow State University Skobeltsyn Institute of Nuclear Physics (MSU SINP), Leninskie gory, Moscow 119991, Russia} 
\maketitle
\begin{history}
\received{25 July 2011}
\end{history}

\begin{abstract}
The single top quark production has an electroweak nature and provides an additional to the top pair production source of the top quarks. The processes involving single top have unique properties, they are very interesting from both theoretical and experimental view points.  
Short review of the single top quark production processes is given in the paper.
\end{abstract}

\ccode{PACS numbers: 11.25.Hf, 123.1K}

\section{Introduction}
The top quark has been discovered at the proton-antiproton 
collider Tevatron in 1995 by two collaborations CDF\cite{top_cdf_1995_ttbar} and D0\cite{top_d0_1995_ttbar}. This discovery was a triumph of the Standard Model (SM) since the top 
quark was found in the mass interval predicted before from a detail 
comparison of LEP measurements  and SM computations obtained at quantum 
loop  level of accuracy\cite{Blondel:1996wm}. More than 10 years later, the top quark was found 
in so-called single top production process which according to SM has a 
cross section only 2.5 times smaller than pair production. The direct single top 
observation  by the Tevatron experiments\cite{Abazov:2009ii,Aaltonen:2009jj} was one more important 
confirmation of our understanding of SM as a quantum gauge field theory describing the Nature 
at extremely small distances of the order of $10^{-17}$ cm.  
Why single top is specially interesting, why it took so long to discover it at the 
Tevatron, how it was discovered at Tevatron and rediscovered at 
the LHC? What may tell us the study of the single top about possible physics 
beyond the Standard Model? This review is aimed to answer these questions 
in some details. 

In SM the top quark is the spin-${1\over 2}$ fermion with 
the electric charge $Q_{em}^t={2\over 3 }\mid e \mid$, the weak isospin 
partner of the $b$ quark, and a color triplet.
Top quark is needed in SM to ensure a cancellation of chiral anomaly\footnote{For the 
cancellation of the anomaly in each generation the sum of electric 
charges of leptons should be equal with the opposite sign to the sum 
quark charges: $(Q_{top}+Q_{b})\times N_c + Q_{/tau} = 0$}
and therefore to ensure a consistency of SM as a quantum theory.
All the couplings and charges of the top 
quark are predicted in SM to be exactly the same as for other two up-type 
quarks, u-quark and c-quark. A natural question one may ask is why then 
the top quark is special and interesting. 

The difference with the other quarks comes from two experimental facts, namely, 
a very large top quark mass comparing to masses of all other quarks and 
very small mixing to quarks of the first and second generations.   
The measured value of the top quark mass is known now with  a precision 
better than 1\%  
$M_{top} = 173.3 \pm  0.6({\rm stat}) \pm  0.9({\rm syst})$ GeV\cite{1007.3178:1900yx}
being the most precisely known quark mass.

The top quark is the heaviest elementary particle found so 
far with a mass slightly less than the mass of the gold nucleus.
In various respects the top quark is a very unique object.
Top Yukawa coupling $\lambda_{t} = 2^{3/4}G_F^{1/2}m_{t}$ is
very close to unit.   
The quark mixing in SM is encoded in matrix elements of the 
Cabibbo-Kobayashi-Maskawa matrix. The matrix element $V_{tb}$ is close 
to one while the elements $V_{ts}$ and $V_{td}$ are significantly smaller 
than one. These two experimental facts, large mass ans small mixing, lead to the conclusion that 
in SM the top quark decays to W-boson and b-quark with a probability close 
to 100 \%. The width of the top quark being calculated in SM at the NLO 
level\cite{Jezabek:1993wk} is about 1.4 GeV. From one side, the top width is much smaller 
than its mass, and therefore the top quark is a narrow resonance 
(top decay width is proportional to the third power of its mass). From the 
other side the top width is significantly larger than a typical QCD scale 
$\Lambda$ 200 MeV. As a result the top quark life time ($\tau_t \approx 5\times 10^{-25}$s) as predicted by SM is 
much smaller than a typical time for formation of QCD bound states ($\tau_{\rm QCD}\approx 
1/\Lambda_{\rm QCD} \approx 3\times 10^{-24}$s).     
Therefore, the $t$-quark decays long before 
it can hadronize and hence top quark containing hadrons do not exist \cite{Bigi:1986jk}. The top quark provides a very clean
source for fundamental information. 

Since the top quark decays before hadronization its spin properties
are not spoiled. Therefore the spin correlation in top  production and decays
is an interesting issue of the top quark physics.

In the single top quark $t$--channel and $s$--channel production processes
the top quark is produced in SM through the left-handed interaction. 
The production is very similar to the top decay process turned backward in time. For the polarized top decay,
it is well-known that the charged lepton tends to point along the direction of top spin\cite{Jezabek:1994zv}.
In the production process this is the direction of the initial
$\bar{d}$-quark for the $s$-channel, and the dominant direction of the
final $d$-quark for the $t$-channel. Therefore, in the top quark rest frame there is strong correlation in the angle of produced lepton with respect to one of the above directions\cite{Mahlon:1996pn,Boos:2002xw}:
\begin{equation}
\frac{1}{\sigma}\frac{d\sigma}{d\cos\theta^*_{\ell}} = \frac{1}{2}(1+\cos\theta^*_{\ell}). 
\label{cos_theta}
\end{equation}
Spin properties in the $tW$ production process
are more involved. Here, one can find a kinematic region in which top
quarks are produced with the polarization vector preferentially close
to the direction of the charged lepton or the $d,s$-quark momentum
from the associated $W$ decay. In this kinematic region, the direction
of the produced charged lepton or the $d,s$-quark should be as close
as possible to the direction of the initial gluon beam in the
top quark rest frame\cite{Boos:2002xw}.

One can also measure several others quantum numbers of the top quark.
One can extract the electric charge 
by measuring the process $t\bar{t}\gamma$ where photon radiates
off the top. The weak isospin of top would be confirmed by looking on
the Wtb vertex structure via top decay in pair production and via
single top production. The confirmation that
the top quark is a color triplet follows from precision measurements 
of the top pair production cross section.

 Since the top is so heavy and point-like at the same time
one might expect a possible deviations from the SM predictions more
likely in the top sector. Top quark physics will be a very important part of research programs
for all future hadron and lepton colliders including studies of top
quark properties, various new physics via the top quark, and
kinematical characteristics of top quark events as significant backgrounds 
to a number of other processes. In particular, the single top production plays a special role here due to its unique properties.
Many details of theoretical studies and experimental analysis of single top production and decay properties could be found in a number of  review papers Ref.~\refcite{Beneke:2000hk}--\refcite{Schilling:2012dx}.

\section{Computation and modeling of processes with single top quark }

\begin{figure}[!h!]
\centerline{\psfig{file=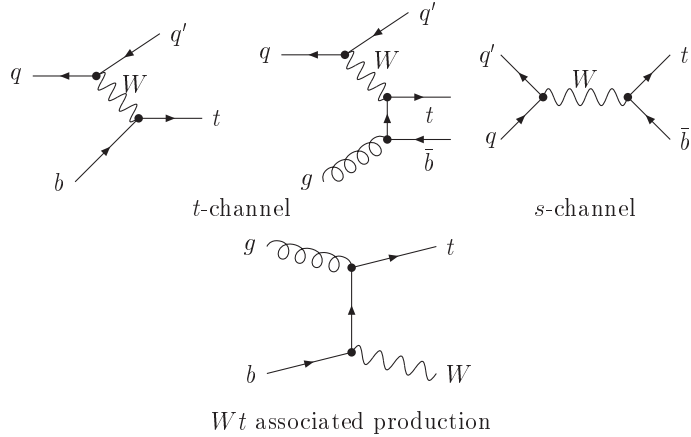,width=10cm}}
\vspace*{8pt}
\caption{The representative diagrams for the single top production 
at the Tevatron and LHC colliders\label{diag_lhc}}
\end{figure}

At hadron and lepton colliders top quarks are produced either in
pairs or singly. 
The representative diagrams for the single top production
at the Tevatron and LHC are shown in Fig.~\ref{diag_lhc}.
Three mechanisms of the single top production are distinguished 
by the virtuality $Q^2_W$ of the W-boson involved:
$t$--channel ($Q^2_W < 0$),
$s$--channel ($Q^2_W > 0$),
associated tW ($Q^2_W = M^2_W$).

The single top quark production at hadron
colliders was considered for the first time in Ref.~\refcite{Willenbrock:1986cr} and later in Ref.~\refcite{Yuan:1989tc}--\refcite{Belyaev:1998dn}. The authors
of Ref.~\refcite{Heinson:1996zm,Belyaev:1998dn} studied the most complete tree-level set of processes
in the SM that contribute to the single top quark production.
QCD NLO corrections to various single top production processes have been calculated in several papers~\refcite{Bordes:1994ki}--\refcite{Kidonakis:2012db}.  In particular, NLO corrections to kinematic distributions were presented\cite{Harris:2002md}. 
The influence of NLO corrections not only to the production but also to the subsequent top quark decay has been studied in Ref.~\refcite{Campbell:2004ch,Schwienhorst:2010je}. Potentially important corrections at the threshold region have been resummed up to NNLL  accuracy\cite{Kidonakis:2010tc,Kidonakis:2010ux,Kidonakis:2011wy}.
Monte-Carlo (MC) analyses of the production processes of
the single top quark allowing to extract it from main backgrounds
were performed in Ref.~\refcite{Belyaev:1998dn,Stelzer:1997ns}.

The NLO cross sections including NNLL soft gluon threshold correction resummation for the main single top production processes at hadron colliders are collected in the Table~\ref{tab:nlocs}:

\begin{table}[!h!]
\tbl{Theoretical cross sections in pb for the main single top production processes at hadron colliders.}
{\begin{tabular}[7]{c|c|c|c}
\hline\hline
&$s$ channel&$t$ channel&$Wt$\\
\hline
Tevatron\cite{Kidonakis:2006bu} ($\sqrt s=1.96$ TeV $p\bar p$)&$1.04\pm 4\%$&$2.26\pm 5\%$&$0.14\pm
20\%$ \\
LHC\cite{Kidonakis:2011wy,Kidonakis:2010tc} ($\sqrt s=7$ TeV $pp$)&$4.6\pm 5\%$&$64\pm 4\%$&$15.6\pm 8\%$\\
LHC\cite{Kidonakis:2012db} ($\sqrt s=8$ TeV $pp$)&$5.55\pm 4\%$&$87.2^{+4}_{-3}\%$&$11.1\pm 7\%$\\
LHC\cite{Kidonakis:2007ej} ($\sqrt s=14$ TeV $pp$)&$12\pm 6\%$&$243\pm 4\%$&$75\pm 10\%$\\
\hline\hline
\end{tabular}\label{tab:nlocs}}
\end{table}

\begin{figure}[!h!]
\centerline{\psfig{file=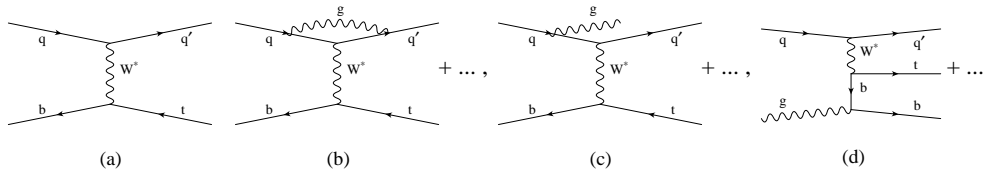,width=13cm}}
\vspace*{8pt}
\caption{The representative NLO diagrams for the $t$--channel single top production 
at the Tevatron and LHC colliders\label{fig:nlo_diag}}
\end{figure}

The processes of the single top-quark production
were simulated using  MC event generators such as ONETOP\cite{Carlson:1993dt} and
TopReX\cite{Slabospitsky:2002ag}, and MC generators based on more generic 
packages such as MadGraph\cite{Herquet:2008zz}, CompHEP\cite{Boos:2004kh},
PYTHIA\cite{Sjostrand:2006za}, AcerMC\cite{Kersevan:2004yg}, MC@NLO\cite{Frixione:2005vw}, and POWHEG\cite{Alioli:2009je}. 
There are several problems associated with the
correct and precise simulation of the single top quark
production processes. Some of these problems
are listed below.
\begin{itemize}
\item The combination of events corresponding to
the diagram in Fig.~\ref{fig:nlo_diag}(a) allowing for the parton showers
in the initial state (ISR) and to the diagrams in
Figs.~\ref{fig:nlo_diag} (b), (c), and (d) results in double counting. Such a double  counting  takes place in
a  soft $P_T$ region of produced b-quark originated from the ISR to the diagram (a) where gluon splits to $b\bar b$ pair and the diagram (d). 
One may subtract the first term in the gluon splitting part to remove the double counting and this procedure gives the correct production cross section. However,
the direct application of the subtraction
procedure for MC event generation for the process results
in a negative weight for part of the events. 
\item Matching procedure for matrix elements and parton showers should be included into event generation.
The modern standards require LHEF format for generated events to be
useful in experimental analyses.
\item As emphasized in Ref.\cite{Mahlon:1996pn,Boos:2002xw}, the top quark is
produced in electroweak processes with significant
polarization owing to the $(V-A)$ structure of the
Wtb vertex in the SM. As a result, spin correlations
between the production and the decay of the top quark
appear. Therefore, the correct MC generator should
include these correlations.
\item  The single top-quark
production processes are sensitive to various new physics contributions\cite{Tait:2000sh} such as anomalous contributions to the Wtb vertex\cite{Kane:1991bg,Boos:1999dd,AguilarSaavedra:2008zc,Zhang:2010dr},
FCNC couplings\cite{Beneke:2000hk} and additional scalar and vector bosons. 
In order to study such extensions of the SM,
 MC generators should include the corresponding anomalous contributions in the production as well in the subsequent top quark decay. 
\item  At the LHC collider, $t$ and $\bar t$ quarks are produced
with different cross sections. The corresponding
asymmetry in the kinematic distributions is useful
for reducing the systematic errors in the measurement
of the top quark parameters\cite{Boos:1999dd}. Therefore, it
is necessary to have the possibility to separate the
production models for $t$ and $\bar t$ quarks at the level of
the MC generator.
\item In case of $tW$-associated production channel one should carefully split the electroweak contribution from QCD top quark pair production\cite{Tait:1999cf,Belyaev:2000me,Campbell:2005bb}. 
\end{itemize}
The mentioned MC event generators try to solve the above problems. However, non of the generators resolve all the problems completely.

\begin{figure}[!h!]
\centerline{\psfig{file=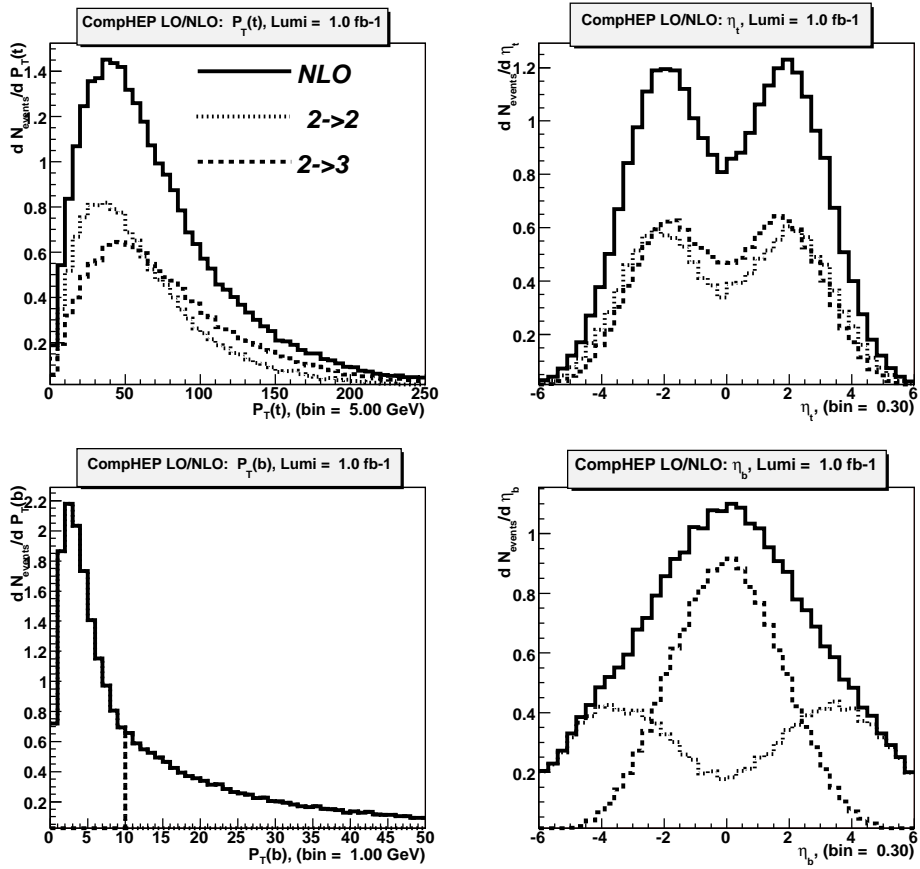,width=13cm}}
\caption{Distributions after the combination of $2\to 2$ events for the $pp\to tq+b_{ISR}$ (ISR is simulated by the PYTHIA generator) and $2\to 3$ events of $pp\to tq+b_{LO}$ (calculated using the CompHEP package) processes 
at the LHC collider with the matching parameter $P_T^0(b)=10$ GeV.
\label{distributions_lhc}}
\end{figure}

In the analysis of the Tevatron data leading to the first observation of the single top production by D0 and CDF collaborations, an effective NLO approach\cite{Boos:2006af} for event generation was used. 
The method was first implemented in the analysis of physics prospects of the CMS experiment and 
described in Ref.~\refcite{CERN-CMS-NOTE-2000-065}.  
The method is realized in the SingleTop MC generator\cite{Boos:2006af} for the analysis in the D0 experiment and in the generator based on MadGraph/MadEvent package\cite{Herquet:2008zz} in the CDF experiment. 
This method of simulation helps in modeling of t- and tW-channels. 
Special analysis has shown that for the s-channel one can simply generate LO events with NLO k-factor. In this case, all of the kinematic distributions obtained from simulated events are in a complete agreement with the same distributions obtained by NLO computations\cite{Sullivan:2004ie}.
The simulation of the $t$-channel process was performed in the five-flavor scheme in which the $2\to 2$ diagram Fig.~\ref{fig:nlo_diag}(a) with the b-quark in the initial state is the leading order contribution. Diagram~\ref{fig:nlo_diag}(b) represents one of the loop NLO contribution while diagrams~\ref{fig:nlo_diag}(c) and~\ref{fig:nlo_diag}(d) represent tree NLO contributions.
One should stress that the diagram~\ref{fig:nlo_diag}(d) gives the NLO $P_T(b)$ spectrum of produced b-quark at high $P_T(b)$ region. One can reproduce low $P_T(b)$ region by switching on ISR corrections to the diagram~\ref{fig:nlo_diag}(a).
All loop and radiation corrections (diagrams~\ref{fig:nlo_diag}(b) and~\ref{fig:nlo_diag}(c)) do not change high $P_T(b)$ spectrum since they are not involved produced b-quark. They affect a renormalization of very soft $P_T(b)$ region and
 therefore can be included numerically by a proper normalization. 
 Such a normalization is performed by summing up hard $P_T(b)$ region as it follows from exact tree NLO computation and soft $P_T(b)$ region multiplying it by some coefficient. This coefficient and phase space slicing parameter in $P_T(b)$ which separates hard and soft regions are determined from two requirements, the first is that the sum of two contributions should be equal to the total NLO cross section and the second is that the $P_T(b)$ distribution should be smooth. In this way, one can combine generated events in the soft and hard regions to one event sample with correct NLO rate and all smooth distributions without negative weights (see examples of such distributions in Fig.~\ref{distributions_lhc}). 

In SingleTop generator, the CompHEP\cite{Boos:2004kh} package is used for complete tree-level computations and NLO rate is calculated by means of the MCFM tool\cite{Campbell:2004ch}. In this way, all spin correlations between production and subsequent top and W boson decays, particle masses and nontrivial decay widths are taken into account. The use of CompHEP allows one to generate events for different extensions of the SM such as mentioned above anomalous Wtb couplings, FCNC couplings and new scalar or vector resonances.

\begin{figure}[!h!]
\centerline{\psfig{file=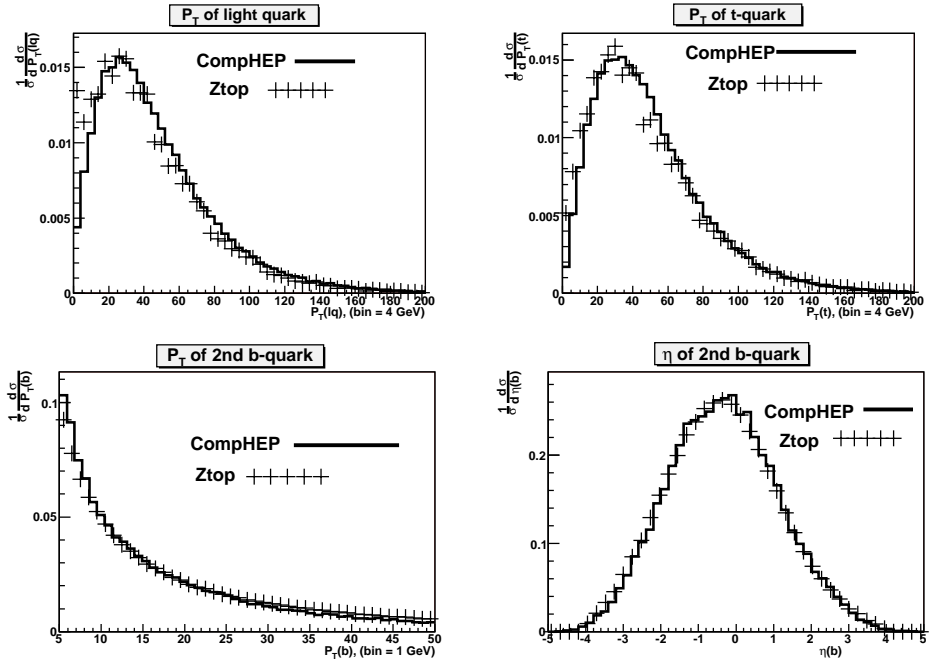,width=13cm}}
\caption{Distributions of the final quarks in the transverse momentum and pseudorapidity in the effective NLO approximation (simulated by the SingleTop generator) and in the exact NLO approximation (obtained by the ZTOP code) for the Tevatron collider. \label{distributions-ztop}}
\end{figure}

\begin{figure}[!h!]
\centerline{\psfig{file=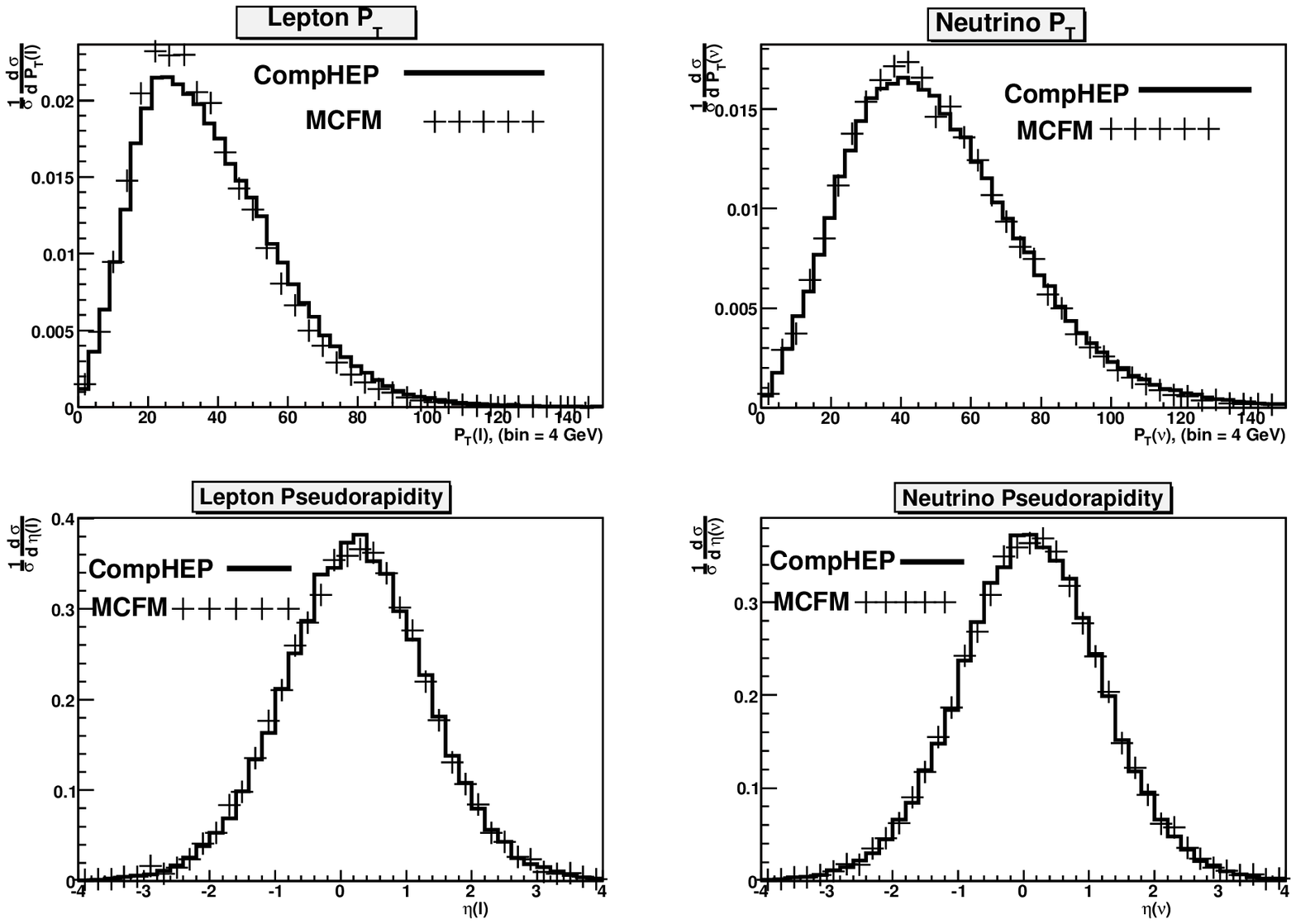,width=13cm}}
\vspace*{8pt}
\caption{Distributions of the transverse momentum and pseudorapidity for the lepton and neutrino from the decay of the top quark in the effective NLO approximation (simulated by the SingleTop generator) and in the exact NLO approximation (calculated by the MCFM code) for the Tevatron collider. \label{distributions-mcfm}}
\end{figure}

Results of a comparison of  various kinematic distributions obtained from the events generated with the SingleTop and computed by means of NLO codes ZTOP\cite{Sullivan:2004ie} and MCFM\cite{Campbell:2004ch}
are shown in Figs.~\ref{distributions-ztop},~\ref{distributions-mcfm} and demonstrate very good agreement.

\begin{figure}[!h!]
\centerline{\psfig{file=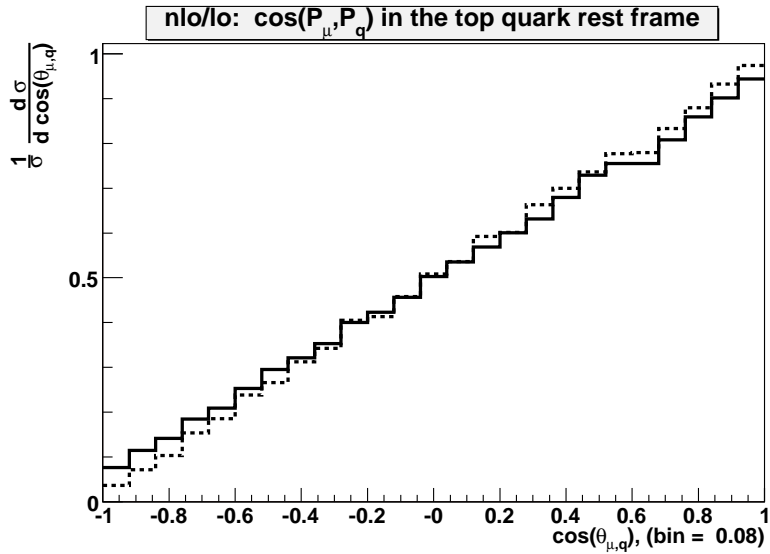,width=11cm}}
\vspace*{8pt}
\caption{ Distribution in the cosine of the angle between
the momenta of the light quark and lepton from the decay
of the top quark in its rest frame. This variable is the 
best to observe spin correlations in
the $t$--channel process. The solid and dotted histograms
correspond to the NLO and LO events, respectively. }
\label{spin_lhc}
\end{figure}

The best spin correlation variable $\cos\theta^*_{\ell}$ (eq.~\ref{cos_theta}) is of a special interest as was mentioned in the introduction, because of unique polarization properties of the single top processes.
Therefore, it is important to show the influence of the NLO corrections to the distribution obtained from the generated events. 
It is easy to show that for the $s$--channel process the spin projection axis corresponding
to the maximum polarization is the momentum
direction of the $\bar{d}$--quark from the initial state in the
rest frame of the top quark\cite{Mahlon:1996pn}. Owing to the correspondence
between the decay and production diagrams
of the top quark (the diagrams are topologically
equivalent), the best probe for the top quark spin is
its decay-product lepton\cite{Boos:2002xw}. Thus, the best variable
to observe spin correlations in the $s$--channel process
is the cosine of the angle between the momenta of
the initial $\bar{d}$--quark and the lepton in the rest frame of
the top quark. Spin correlations can be numerically
characterized by the coefficient $R_{\rm spin}$ of $\cos{\theta^*_{e^+,\bar{d}}}$ in the
normalized distribution
$$\frac{1}{\sigma}\frac{d\sigma}{d\cos{\theta^*_{e^+,\bar{d}}}} = 
\frac{1+R_{spin}(\bar{s})\cos{\theta^*_{e^+,\bar{d}}}}2.$$
Then, $R_{\rm spin}(\bar{p}_d)=1$ (or 100\%) for the $s$--channel process.
Since the NLO approximation is manifested
only in the $K$--factor in this process, we do not expect
any significant reduction of $R_{\rm spin}$ owing to the inclusion
of NLO corrections.
The diagram of the $t$--channel process in the LO
approximation is also topologically equivalent to the
decay diagrams of the $s$--channel process. Thus, the
top quark is polarized, and the axis of the maximum
polarization is the momentum of the final light
quark in the rest frame of the top quark. The dotted
histogram in Fig.~\ref{spin_lhc} corresponds to LO events.
The first-order polynomial fit to the distribution gives
$R_{\rm spin}(\bar{p}_d)_{\rm LO}=0.98\pm0.02$, which indicates the maximum
polarization of the top quark in the LO approximation.
In the NLO approximation, a significant
contribution comes from the real correction with the
additional b-quark. In this process, the top quark can
be produced in the QCD vector interaction vertex
with the gluon, which reduces the polarization of
the top quark. However, this reduction is not strong
because the main contribution to the $pp\to tqb$ process
comes from the diagram with the Wtb production
vertex of the top quark. The solid histogram in
Fig.~\ref{spin_lhc} shows the distribution of the NLO events in $\cos{\theta^*_{l^+,d}}$
The straight-line fit of the distribution gives
$R_{\rm spin}(\bar{p}_d)_{\rm NLO}=0.89\pm0.02$, which indicates the
reduction of the polarization value\cite{Boos:2006af}.

\begin{figure}[!h!]
\centerline{\psfig{file=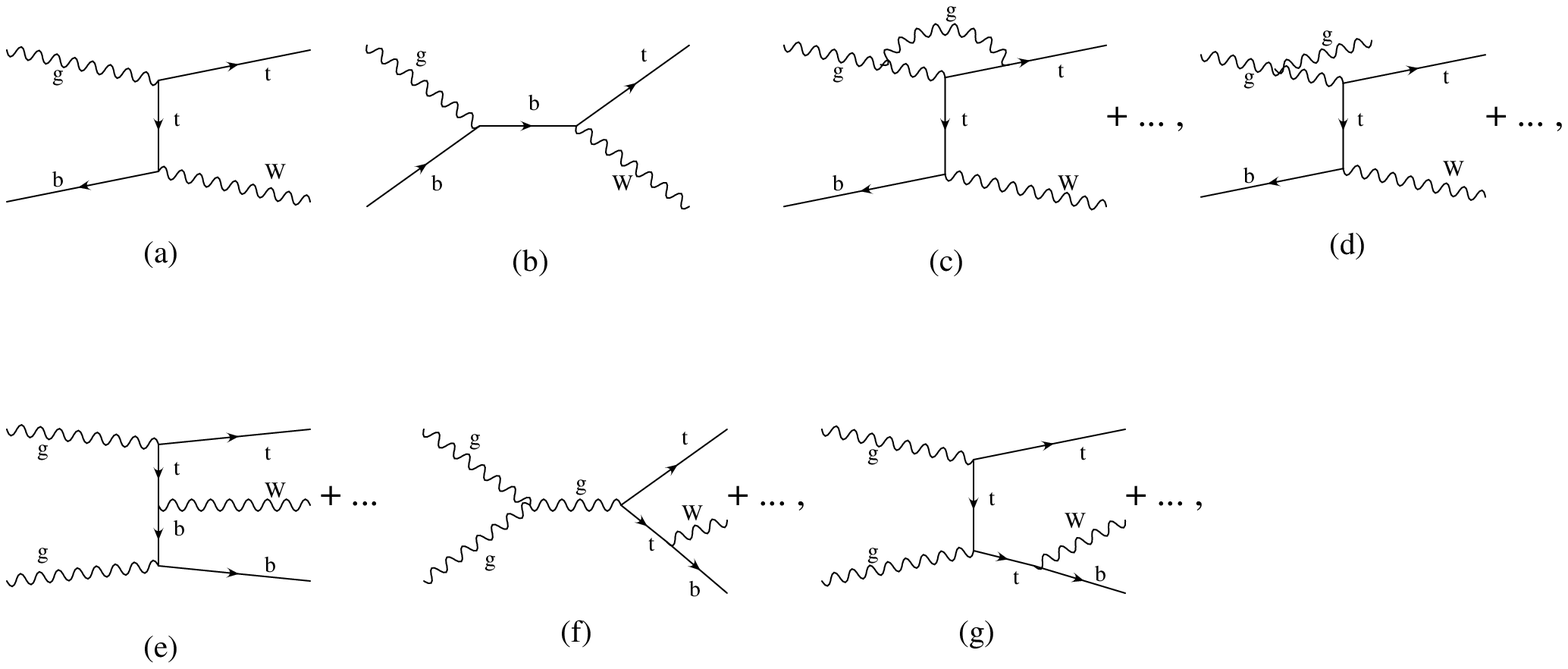,width=15cm}}
\vspace*{8pt}
\caption{The representative LO and NLO diagrams for the $tW$--channel single top production 
at the Tevatron and LHC colliders\label{fig:tw_diag}}
\end{figure}

The $tW$--process requires special consideration.
The LO diagrams are shown on the Fig.~\ref{fig:tw_diag}(a,b). The diagrams (c,d) are the representative diagrams for NLO loop and tree corrections involving gluons and light quarks. Diagram (e) is one of 
the tree NLO contribution with additional b-quark produced, similar to the diagram in Fig.~\ref{fig:nlo_diag}(d) for the t-channel process.
Diagrams~\ref{fig:tw_diag}(f) and~\ref{fig:tw_diag}(g) contain the top pair contribution with a subsequent decay of the second top to $W$ and b. There are many other subleading diagrams mentioned in Ref.~\refcite{Belyaev:2000me} making the result exactly electroweak gauge invariant. The problem of double counting of contributions of diagrams (a,b) and (e) is very similar to that discussed here for t-channel case and is resolved in the same manner by slicing phase space on hard and soft $P_T(b)$ regions. However, in $tW$ case there is another problem of how to split single top and  top pair contributions leading to the same final states. In order to do that, several methods have been proposed. In the paper\cite{Tait:1999cf}, the top pair part was removed by subtraction top pairs on shell. This procedure called the diagram subtraction scheme is obviously gauge invariant and later was realised on the generator level in MC@NLO code\cite{Frixione:2005vw}.
The procedure includes all the interferences of single and top pair contributions into the single top part 
leading, however, to negative weights for some fraction of generated events. In another scheme called the diagram removal scheme also implemented in MC@NLO, diagrams Fig.~\ref{fig:tw_diag}(f,g)  are removed from the complete set of the diagrams. In this approach, all the interferences between single and top pair are removed leading, however, to a small violation of the electroweak gauge invariance. In the paper\cite{Belyaev:2000me}, the phase space removal scheme was used in which the part of events with the $Wb$ invariant mass around the top quark pole was removed. In this way all the interferences are removed, however, there is an ambiguity in the size of the removal $Wb$ invariant mass region. In the paper\cite{Campbell:2005bb}, an approach with veto on the $P_T(b)$ larger than some $P_T^{\rm veto}(b)$ separation parameter was used. In this approach, some small part of top pair contribution might still belong to single top part and also there is some ambiguity in choosing of the separation 
parameter. 
However, in practice the veto is realisable only in some part of the phase space region where the b-quark could be observed as a b-jet. 

Spin correlations in $tW$-channel requires special consideration. 
The leading order Feynman diagrams for this process
include two diagrams: in Fig.~\ref{fig:tw_diag}(a), the top quark
is produced in the QCD interaction vertex, and in another one Fig.~\ref{fig:tw_diag}(b) it is produced in the electroweak interaction vertex.
Both contributions are comparable in rate. The first diagram leads to unpolarized top quark production while from the second diagram the top is produced highly polarized. If the top quark would be produced only from electroweak diagram Fig.~\ref{fig:tw_diag}(b), its spin direction in the rest frame would fully correlated with a direction of charged lepton momentum coming from the $W$ boson decay. However, this property is spoiled by the contribution of the first diagram Fig.~\ref{fig:tw_diag}(a). As was shown in\cite{Boos:2002xw},
one can apply some kinematic cuts in order to suppress the contribution of the diagram Fig.~\ref{fig:tw_diag}(a) making  the spin correlation property more pronounced. In this region, the outgoing charged lepton should be as close as possible to the direction of one of the initial beam in the top quark rest frame.
It was shown that one can increase the
polarization of the top quark from the initial 24 (without any cuts) to
80 -- 90\% applying some reasonable cuts.

\section{Single top observation at the Tevatron}

Two years before the top quark was discovered, D0 collaboration had organized single top research group to perform a search for the single electroweak top quark production. The cross section of the single top processes is only about two times smaller than the top pair production, but the number of final jets is smaller and 
therefore the backgrounds are significantly higher. During the Run I data taking, D0 detector did not have vertex detector and could not identify b-quarks with high efficiency (CDF detector had the part). Since the top quark decays almost 100\% with production of b-quark, this feature of the detector is very important in top physics. The small ratio of signal to backgrounds and lack of the vertex detector were the main reasons to find sophisticated methods to increase signal to background ratio and achieve sensible result. It was a powerful stimulation for the analyzers to implement and develop multivariate analysis techniques. The main strategy in D0 analysis was to apply loose initial 
selection to keep the signal statistics and apply multivariate analysis technique to distinguish the signal events. CDF strategy included rather tight initial selection to cut the contribution of pair production by the cuts on number of jets (only two jet events) and number of $b$-quarks (only one $b$-quark in event). Due to the complexity of the analysis and lack of experimental statistics the first results\cite{Abazov:2001ns} were published in about 7 years after top quark was discovered. The available statistics in Run I analysis did not allow to observe the single top processes and both collaborations set upper limits on the production cross section. Because of multivariate technique in the analysis, D0 collaboration succeeded in achieving the same sensitivity as CDF collaboration with the use of vertex detector. Both the collaborations set 95\% C.L. upper limits on the production cross section for s-channel process 17 pb is the D0 results and 18 pb is the CDF result; for the t-channel process 13 pb is the CDF 
result and 22 pb is the D0 result. 

During the second run of the Tevatron (Run II), the luminosity has increased significantly and collaborations upgraded their detectors. The first evidence of the single top quark production was reported by D0 collaboration in December of 2006\cite{Abazov:2006gd} and later by CDF collaboration\cite{Aaltonen:2008sy}, with the first measurement of the production cross section and the first direct measurement of the $V_{tb}$ CKM matrix element. After the luminosity has reached 2-3 $fb^{-1}$ both of the collaborations reported $5\sigma$ observation of the single top production\cite{Abazov:2009ii}. Based on the full Run II statistics, both collaborations are able to distinguish $s-$ and $t$-channel processes and measure their cross sections separately. This measurement significantly increases the sensitivity for the possible BSM contribution. The current results of the SM measurements in D0\cite{Abazov:2011pt} with integrated luminosity of 5.4 $fb^{-1}$ are the $\sigma({\ppbar}{\rargap}tb+X, tqb+X) = 3.43\pm^{0.73}_{0.74}\rm pb$ the corresponding measurement of the CKM matrix element is $0.79 < |V_{tb}| \leq 1$ at the 95\% C.L. The separate measurements of $s-$ and $t$-channel are the following $\sigma({\ppbar}{\rargap}tb+X) = 0.68\pm^{0.38}_{0.35}\rm pb$ and $\sigma({\ppbar}{\rargap}tqb+X) = 2.86\pm^{0.69}_{0.63} \rm pb$. The current results of the SM measurements in CDF\cite{Aaltonen:2010jr} with integrated luminosity of 3.2 $fb^{-1}$ and $5\sigma$ statistical significance are $\sigma({\ppbar}{\rargap}tb+X, tqb+X) = 2.3\pm^{0.6}_{0.5}$(stat+sys) pb, the measured CKM matrix element value $|Vtb|=0.91\pm^{0.11}_{0.11} {\rm (stat+sys)} \pm 0.07$(theory) with a lower 95\% C.L. limit $0.71<|Vtb|$.

\section{Single top evidence and observation at the LHC}

The relative contribution of different single top production channels significantly differs for LHC than at the Tevatron. All the processes with initial gluon are significantly larger than the contribution of the processes with initial quarks. Therefore, the main mechanisms of single top production at the LHC are $t$- and $tW$-channels, but the $s$-channel cross section is significantly lower (Table~\ref{tab:nlocs}).  High luminosity and relatively high cross section result in possible very high statistics of single top events at the LHC and the main limitation for the measurements is the systematic uncertainty. During the first round of analysis at $\sqrt{s}=7$ TeV, CMS and ATLAS collaborations measured cross section of the $t-$ and $tW-$channel processes
and set the first upper limits for $s$-channel production cross section. 
CMS collaboration reported\cite{Chatrchyan:2011vp} the first evidence of $t$-channel single top production at the LHC and measure the cross section $83.6 \pm 29.8 {\rm (stat.+syst.)} \pm 3.3 $ (lumi.) pb 
with 3.5$\sigma$ significance at 36 $\rm pb^{-1}$ of integrated luminosity and 95\% C.L. limit for CKM matrix element $0.68<|Vtb|$. This result was improved with the higher statistics\cite{CMS-t-channel}:  $70.2 \pm 5.2 {\rm (stat.)} \pm 10.4 {\rm (syst.)}\pm 3.4 $ (lumi.) pb. 
CMS collaboration made the first measurement of the $tW$-channel cross section\cite{CMS-tw}$\sigma(tW)=16\pm^{5}_{4}{\rm (stat.+syst.)}$ with 4$\sigma$ of the observed significance with 4.9 $fb^{-1}$ of integrated luminosity. The ATLAS collaboration has reported\cite{ATLAS-t} observation of $t$-channel single top production with cross section $90\pm 9 {\rm (stat.)} \pm^{31}_{20} {\rm (syst.)}$ pb at 7.6$\sigma$ significance with 0.7 $\rm fb^{-1}$ of integrated luminosity.  The first evidence of $tW$-channel production (at 2.05 $\rm fb^{-1}$) has reported by ATLAS\cite{:2012dj} with $3.3\sigma$ significance and measured cross section $\sigma(tW) = 16.8 \pm 2.9  {\rm (stat)} \pm 4.9  {\rm (syst)}$ pb, translated to measurement of   $|V_{tb}| = 1.03^{+0.16}_{-0.19}$.
ATLAS has set the first LHC $s-$channel limits\cite{ATLAS-s} $\sigma(s-{\rm channel})<26 pb$. Both of the collaborations have started the searches for the ``New Physics'' effects in single top production.

\section{Single top at future linear colliders}

For completeness we discuss in this section single top quark production at future lepton colliders. In $e^+e^-$ collisions, the top quarks
can be produced in pairs or singly similarly to hadronic collisions. However, an important difference is that at lepton collider both pair and single production processes have the same electroweak origin while at hadron colliders the top quark pair is a strong production process. Due to the fact of the electroweak nature, both processes may simultaneously give contributions to the same final states. Therefore, a special care should be paid to split these two contributions correctly.   
 
To illustrate this, let us consider for simplicity the case when one of the top remains stable.  In this case, the contributing diagrams for both pair and single top quark production
in $e^+e^-$ collisions are shown in Figs.~\ref{eetops} and \ref{eetopt} for
 $e \nu_{e} b t$ final state\cite{Boos:2001sj}. The diagrams form so-called CC20\cite{Bardin} set of diagrams which
splits into two gauge invariant subsets of 10 diagrams, s-channel and t-channel (see\cite{boos-ohl}. The s-channel subset contains only two diagrams (diagrams 1,2 in Fig.~\ref{eetops}) with top pair production and subsequent decay. All other diagrams in s-channel and t-channel subsets contribute to the single top production. The other possible final states correspond to the top
pair production and follow from possible decay modes of $W$-boson 
$e^+e^- \to t\bar{t} \to W W b \bar{b}, ~~~~~W \to f\bar{f^{\prime}}$,\\
where e.g. for $W^+$
$f = u,c,\nu_{e},\nu_{\mu},\nu_{\tau}\nu_{\mu}$;
      $f^{\prime} = d,s,e,\mu,\tau$

\begin{figure}[!h!]
{
\unitlength=0.5 pt
\SetScale{0.5}
\SetWidth{0.7}      
\scriptsize    
\noindent
{} \allowbreak
\begin{picture}(95,99)(0,0)
\Text(15.0,90.0)[r]{$e$}
\ArrowLine(16.0,90.0)(37.0,80.0) 
\Text(15.0,70.0)[r]{$\bar{e}$}
\ArrowLine(37.0,80.0)(16.0,70.0) 
\Text(47.0,81.0)[b]{$\gamma,Z$}
\DashLine(37.0,80.0)(58.0,80.0){3.0} 
\Text(80.0,90.0)[l]{$t$}
\ArrowLine(58.0,80.0)(79.0,90.0) 
\Text(54.0,70.0)[r]{$t$}
\ArrowLine(58.0,60.0)(58.0,80.0) 
\Text(80.0,70.0)[l]{$\bar{b}$}
\ArrowLine(79.0,70.0)(58.0,60.0) 
\Text(54.0,50.0)[r]{$W^+$}
\DashArrowLine(58.0,40.0)(58.0,60.0){3.0} 
\Text(80.0,50.0)[l]{$e$}
\ArrowLine(58.0,40.0)(79.0,50.0) 
\Text(80.0,30.0)[l]{$\bar{\nu}_e$}
\ArrowLine(79.0,30.0)(58.0,40.0) 
\Text(47,0)[b] {diagr.1,2}
\end{picture} \ 
{} \qquad\allowbreak
\begin{picture}(95,99)(0,0)
\Text(15.0,90.0)[r]{$e$}
\ArrowLine(16.0,90.0)(37.0,80.0) 
\Text(15.0,70.0)[r]{$\bar{e}$}
\ArrowLine(37.0,80.0)(16.0,70.0) 
\Text(47.0,81.0)[b]{$\gamma,Z$}
\DashLine(37.0,80.0)(58.0,80.0){3.0} 
\Text(80.0,90.0)[l]{$e$}
\ArrowLine(58.0,80.0)(79.0,90.0) 
\Text(54.0,70.0)[r]{$e$}
\ArrowLine(58.0,60.0)(58.0,80.0) 
\Text(80.0,70.0)[l]{$\bar{\nu}_e$}
\ArrowLine(79.0,70.0)(58.0,60.0) 
\Text(54.0,50.0)[r]{$W^+$}
\DashArrowLine(58.0,60.0)(58.0,40.0){3.0} 
\Text(80.0,50.0)[l]{$\bar{b}$}
\ArrowLine(79.0,50.0)(58.0,40.0) 
\Text(80.0,30.0)[l]{$t$}
\ArrowLine(58.0,40.0)(79.0,30.0) 
\Text(47,0)[b] {diagr.3,4}
\end{picture} \ 
{} \qquad\allowbreak
\begin{picture}(95,99)(0,0)
\Text(15.0,90.0)[r]{$e$}
\ArrowLine(16.0,90.0)(37.0,80.0) 
\Text(15.0,70.0)[r]{$\bar{e}$}
\ArrowLine(37.0,80.0)(16.0,70.0) 
\Text(47.0,81.0)[b]{$\gamma,Z$}
\DashLine(37.0,80.0)(58.0,80.0){3.0} 
\Text(80.0,90.0)[l]{$\bar{b}$}
\ArrowLine(79.0,90.0)(58.0,80.0) 
\Text(54.0,70.0)[r]{$b$}
\ArrowLine(58.0,80.0)(58.0,60.0) 
\Text(80.0,70.0)[l]{$t$}
\ArrowLine(58.0,60.0)(79.0,70.0) 
\Text(54.0,50.0)[r]{$W^+$}
\DashArrowLine(58.0,40.0)(58.0,60.0){3.0} 
\Text(80.0,50.0)[l]{$e$}
\ArrowLine(58.0,40.0)(79.0,50.0) 
\Text(80.0,30.0)[l]{$\bar{\nu}_e$}
\ArrowLine(79.0,30.0)(58.0,40.0) 
\Text(47,0)[b] {diagr.5,6}
\end{picture} \ 
{} \qquad\allowbreak
\begin{picture}(95,99)(0,0)
\Text(15.0,70.0)[r]{$e$}
\ArrowLine(16.0,70.0)(37.0,60.0) 
\Text(15.0,50.0)[r]{$\bar{e}$}
\ArrowLine(37.0,60.0)(16.0,50.0) 
\Text(37.0,62.0)[lb]{$\gamma,Z$}
\DashLine(37.0,60.0)(58.0,60.0){3.0} 
\Text(54.0,70.0)[r]{$$}
\DashArrowLine(58.0,80.0)(58.0,60.0){3.0} 
\Text(80.0,90.0)[l]{$e$}
\ArrowLine(58.0,80.0)(79.0,90.0) 
\Text(80.0,70.0)[l]{$\bar{\nu}_e$}
\ArrowLine(79.0,70.0)(58.0,80.0) 
\Text(54.0,50.0)[r]{$W^+$}
\DashArrowLine(58.0,60.0)(58.0,40.0){3.0} 
\Text(80.0,50.0)[l]{$\bar{b}$}
\ArrowLine(79.0,50.0)(58.0,40.0) 
\Text(80.0,30.0)[l]{$t$}
\ArrowLine(58.0,40.0)(79.0,30.0) 
\Text(47,0)[b] {diagr.7,8}
\end{picture} \ 
{} \qquad\allowbreak
\begin{picture}(95,99)(0,0)
\Text(15.0,80.0)[r]{$e$}
\ArrowLine(16.0,80.0)(37.0,80.0) 
\Text(47.0,84.0)[b]{$W^+$}
\DashArrowLine(58.0,80.0)(37.0,80.0){3.0} 
\Text(80.0,90.0)[l]{$e$}
\ArrowLine(58.0,80.0)(79.0,90.0) 
\Text(80.0,70.0)[l]{$\bar{\nu}_e$}
\ArrowLine(79.0,70.0)(58.0,80.0) 
\Text(33.0,60.0)[r]{$\nu_e$}
\ArrowLine(37.0,80.0)(37.0,40.0) 
\Text(15.0,40.0)[r]{$\bar{e}$}
\ArrowLine(37.0,40.0)(16.0,40.0) 
\Text(47.0,44.0)[b]{$W^+$}
\DashArrowLine(37.0,40.0)(58.0,40.0){3.0} 
\Text(80.0,50.0)[l]{$\bar{b}$}
\ArrowLine(79.0,50.0)(58.0,40.0) 
\Text(80.0,30.0)[l]{$t$}
\ArrowLine(58.0,40.0)(79.0,30.0) 
\Text(47,0)[b] {diagr.9}
\end{picture} \ 
{} \qquad\allowbreak
\begin{picture}(95,99)(0,0)
\Text(15.0,90.0)[r]{$e$}
\ArrowLine(16.0,90.0)(37.0,80.0) 
\Text(15.0,70.0)[r]{$\bar{e}$}
\ArrowLine(37.0,80.0)(16.0,70.0) 
\Text(47.0,81.0)[b]{$Z$}
\DashLine(37.0,80.0)(58.0,80.0){3.0} 
\Text(80.0,90.0)[l]{$\bar{\nu}_e$}
\ArrowLine(79.0,90.0)(58.0,80.0) 
\Text(54.0,70.0)[r]{$\nu_e$}
\ArrowLine(58.0,80.0)(58.0,60.0) 
\Text(80.0,70.0)[l]{$e$}
\ArrowLine(58.0,60.0)(79.0,70.0) 
\Text(54.0,50.0)[r]{$W^+$}
\DashArrowLine(58.0,60.0)(58.0,40.0){3.0} 
\Text(80.0,50.0)[l]{$\bar{b}$}
\ArrowLine(79.0,50.0)(58.0,40.0) 
\Text(80.0,30.0)[l]{$t$}
\ArrowLine(58.0,40.0)(79.0,30.0) 
\Text(47,0)[b] {diagr.10}
\end{picture} \ 
}
\caption{Diagrams for s-channel top quark production in $e^+e^-$ collisions}
\label{eetops}
\end{figure}
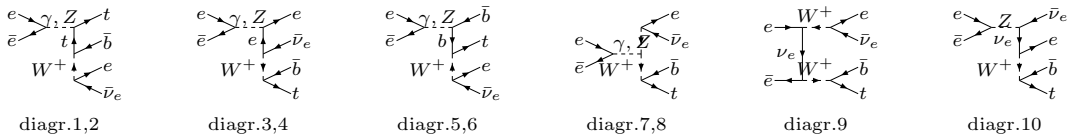
\begin{figure}[!h!]
{
\unitlength=0.5 pt
\SetScale{0.5}
\SetWidth{0.7}      
\scriptsize    
\noindent
{} \allowbreak
\begin{picture}(95,99)(0,0)
\Text(15.0,80.0)[r]{$e$}
\ArrowLine(16.0,80.0)(37.0,80.0) 
\Line(37.0,80.0)(58.0,80.0) 
\Text(80.0,90.0)[l]{$e$}
\ArrowLine(58.0,80.0)(79.0,90.0) 
\Text(36.0,70.0)[r]{$\gamma,Z$}
\DashLine(37.0,80.0)(37.0,60.0){3.0} 
\Text(15.0,60.0)[r]{$\bar{e}$}
\ArrowLine(37.0,60.0)(16.0,60.0) 
\Text(47.0,64.0)[b]{$e$}
\ArrowLine(58.0,60.0)(37.0,60.0) 
\Text(80.0,70.0)[l]{$\bar{\nu}_e$}
\ArrowLine(79.0,70.0)(58.0,60.0) 
\Text(54.0,50.0)[r]{$W^+$}
\DashArrowLine(58.0,60.0)(58.0,40.0){3.0} 
\Text(80.0,50.0)[l]{$\bar{b}$}
\ArrowLine(79.0,50.0)(58.0,40.0) 
\Text(80.0,30.0)[l]{$t$}
\ArrowLine(58.0,40.0)(79.0,30.0) 
\Text(47,0)[b] {diagr.1,2}
\end{picture} \ 
{} \qquad\allowbreak
\begin{picture}(95,99)(0,0)
\Text(15.0,90.0)[r]{$e$}
\ArrowLine(16.0,90.0)(58.0,90.0) 
\Text(80.0,90.0)[l]{$e$}
\ArrowLine(58.0,90.0)(79.0,90.0) 
\Text(57.0,80.0)[r]{$\gamma,Z$}
\DashLine(58.0,90.0)(58.0,70.0){3.0} 
\Text(80.0,70.0)[l]{$t$}
\ArrowLine(58.0,70.0)(79.0,70.0) 
\Text(54.0,60.0)[r]{$t$}
\ArrowLine(58.0,50.0)(58.0,70.0) 
\Text(80.0,50.0)[l]{$\bar{b}$}
\ArrowLine(79.0,50.0)(58.0,50.0) 
\Text(54.0,40.0)[r]{$W^+$}
\DashArrowLine(58.0,30.0)(58.0,50.0){3.0} 
\Text(15.0,30.0)[r]{$\bar{e}$}
\ArrowLine(58.0,30.0)(16.0,30.0) 
\Text(80.0,30.0)[l]{$\bar{\nu}_e$}
\ArrowLine(79.0,30.0)(58.0,30.0) 
\Text(47,0)[b] {diagr.3,4}
\end{picture} \ 
{} \qquad\allowbreak
\begin{picture}(95,99)(0,0)
\Text(15.0,90.0)[r]{$e$}
\ArrowLine(16.0,90.0)(58.0,90.0) 
\Text(80.0,90.0)[l]{$e$}
\ArrowLine(58.0,90.0)(79.0,90.0) 
\Text(57.0,80.0)[r]{$\gamma,Z$}
\DashLine(58.0,90.0)(58.0,70.0){3.0} 
\Text(80.0,70.0)[l]{$\bar{b}$}
\ArrowLine(79.0,70.0)(58.0,70.0) 
\Text(54.0,60.0)[r]{$b$}
\ArrowLine(58.0,70.0)(58.0,50.0) 
\Text(80.0,50.0)[l]{$t$}
\ArrowLine(58.0,50.0)(79.0,50.0) 
\Text(54.0,40.0)[r]{$W^+$}
\DashArrowLine(58.0,30.0)(58.0,50.0){3.0} 
\Text(15.0,30.0)[r]{$\bar{e}$}
\ArrowLine(58.0,30.0)(16.0,30.0) 
\Text(80.0,30.0)[l]{$\bar{\nu}_e$}
\ArrowLine(79.0,30.0)(58.0,30.0) 
\Text(47,0)[b] {diagr.5,6}
\end{picture} \ 
{} \qquad\allowbreak
\begin{picture}(95,99)(0,0)
\Text(15.0,80.0)[r]{$e$}
\ArrowLine(16.0,80.0)(37.0,80.0) 
\Line(37.0,80.0)(58.0,80.0) 
\Text(80.0,90.0)[l]{$e$}
\ArrowLine(58.0,80.0)(79.0,90.0) 
\Text(36.0,70.0)[r]{$\gamma,Z$}
\DashLine(37.0,80.0)(37.0,60.0){3.0} 
\Text(47.0,64.0)[b]{$W^+$}
\DashArrowLine(37.0,60.0)(58.0,60.0){3.0} 
\Text(80.0,70.0)[l]{$\bar{b}$}
\ArrowLine(79.0,70.0)(58.0,60.0) 
\Text(80.0,50.0)[l]{$t$}
\ArrowLine(58.0,60.0)(79.0,50.0) 
\Text(33.0,50.0)[r]{$W^+$}
\DashArrowLine(37.0,40.0)(37.0,60.0){3.0} 
\Text(15.0,40.0)[r]{$\bar{e}$}
\ArrowLine(37.0,40.0)(16.0,40.0) 
\Line(37.0,40.0)(58.0,40.0) 
\Text(80.0,30.0)[l]{$\bar{\nu}_e$}
\ArrowLine(79.0,30.0)(58.0,40.0) 
\Text(47,0)[b] {diagr.7,8}
\end{picture} \ 
{} \qquad\allowbreak
\begin{picture}(95,99)(0,0)
\Text(15.0,80.0)[r]{$e$}
\ArrowLine(16.0,80.0)(37.0,80.0) 
\Text(47.0,84.0)[b]{$\nu_e$}
\ArrowLine(37.0,80.0)(58.0,80.0) 
\Text(80.0,90.0)[l]{$e$}
\ArrowLine(58.0,80.0)(79.0,90.0) 
\Text(54.0,70.0)[r]{$W^+$}
\DashArrowLine(58.0,80.0)(58.0,60.0){3.0} 
\Text(80.0,70.0)[l]{$\bar{b}$}
\ArrowLine(79.0,70.0)(58.0,60.0) 
\Text(80.0,50.0)[l]{$t$}
\ArrowLine(58.0,60.0)(79.0,50.0) 
\Text(33.0,60.0)[r]{$W^+$}
\DashArrowLine(37.0,40.0)(37.0,80.0){3.0} 
\Text(15.0,40.0)[r]{$\bar{e}$}
\ArrowLine(37.0,40.0)(16.0,40.0) 
\Line(37.0,40.0)(58.0,40.0) 
\Text(80.0,30.0)[l]{$\bar{\nu}_e$}
\ArrowLine(79.0,30.0)(58.0,40.0) 
\Text(47,0)[b] {diagr.9}
\end{picture} \ 
{} \qquad\allowbreak
\begin{picture}(95,99)(0,0)
\Text(15.0,80.0)[r]{$e$}
\ArrowLine(16.0,80.0)(37.0,80.0) 
\Line(37.0,80.0)(58.0,80.0) 
\Text(80.0,90.0)[l]{$e$}
\ArrowLine(58.0,80.0)(79.0,90.0) 
\Text(36.0,70.0)[r]{$Z$}
\DashLine(37.0,80.0)(37.0,60.0){3.0} 
\Line(37.0,60.0)(58.0,60.0) 
\Text(80.0,70.0)[l]{$\bar{\nu}_e$}
\ArrowLine(79.0,70.0)(58.0,60.0) 
\Text(33.0,50.0)[r]{$\nu_e$}
\ArrowLine(37.0,60.0)(37.0,40.0) 
\Text(15.0,40.0)[r]{$\bar{e}$}
\ArrowLine(37.0,40.0)(16.0,40.0) 
\Text(47.0,44.0)[b]{$W^+$}
\DashArrowLine(37.0,40.0)(58.0,40.0){3.0} 
\Text(80.0,50.0)[l]{$\bar{b}$}
\ArrowLine(79.0,50.0)(58.0,40.0) 
\Text(80.0,30.0)[l]{$t$}
\ArrowLine(58.0,40.0)(79.0,30.0) 
\Text(47,0)[b] {diagr.10}
\end{picture} \ 
}
\caption{Diagrams for t-channel top quark production in $e^+e^-$ collisions}
\label{eetopt}
\end{figure}
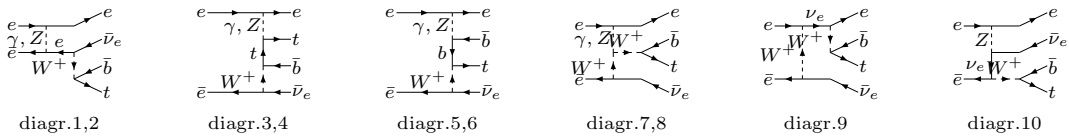

One can compute the single top cross section as the difference 
of the complete tree-level (CTL) contribution and the Breit-Wigner (BW) 
resonance contribution obtained from a proper fit of the invariant mass distribution 
computed from the gauge invariant complete set of tree-level diagrams:
$\int dM_{e \nu b} \, (d\sigma^{CTL}/dM_{e \nu b}-d\sigma^{BW}/dM_{e \nu 
b})$.  One can also use a cut on the $e \nu b$ invariant mass
around the top quark pole\cite{sensitivity,Belyaev:2000me} as an equivalent of 
the BW subtraction procedure
\begin{equation}
\sigma=\int\limits_{M_{min}}^{m_{top}-\Delta} dM_{e \nu b} 
\frac{d\sigma^{CTL}}{dM_{e \nu b}}
      +\int\limits^{M_{max}}_{m_{top}+\Delta} dM_{e \nu b} 
\frac{d\sigma^{CTL}}{dM_{e \nu b}},
\end{equation}
where the value of $\Delta$ is taken to be 20 GeV. With such a value cross sections in both ways of computation agree very well.
 This value of $\Delta$ is much larger 
than  
an intuitively expected one of the 
order of the top quark width, which would lead to large contributions of 
surviving $t \bar t$ events. Obviously, the procedure applied is gauge 
invariant.

\begin{figure}[!h!]
{
\fontsize{3}{6pt}\selectfont
\unitlength=0.46 pt
\SetScale{0.46}
\SetWidth{0.7}      
\tiny\tt    
\begin{picture}(95,79)(0,0)
\Text(15.0,60.0)[r]{$\gamma$}
\DashLine(16.0,60.0)(37.0,60.0){3.0} 
\Text(47.0,64.0)[b]{$t$}
\ArrowLine(58.0,60.0)(37.0,60.0) 
\Text(80.0,70.0)[l]{$W^-$}
\DashArrowLine(79.0,70.0)(58.0,60.0){3.0} 
\Text(80.0,50.0)[l]{$\bar{b}$}
\ArrowLine(79.0,50.0)(58.0,60.0) 
\Text(33.0,50.0)[r]{$t$}
\ArrowLine(37.0,60.0)(37.0,40.0) 
\Text(15.0,40.0)[r]{$\gamma$}
\DashLine(16.0,40.0)(37.0,40.0){3.0} 
\Line(37.0,40.0)(58.0,40.0) 
\Text(80.0,30.0)[l]{$t$}
\ArrowLine(58.0,40.0)(79.0,30.0) 
\Text(47,0)[b] {diagr.1}
\end{picture} \ 
\begin{picture}(95,79)(0,0)
\Text(15.0,60.0)[r]{$\gamma$}
\DashLine(16.0,60.0)(37.0,60.0){3.0} 
\Line(37.0,60.0)(58.0,60.0) 
\Text(80.0,70.0)[l]{$t$}
\ArrowLine(58.0,60.0)(79.0,70.0) 
\Text(33.0,50.0)[r]{$t$}
\ArrowLine(37.0,40.0)(37.0,60.0) 
\Text(15.0,40.0)[r]{$\gamma$}
\DashLine(16.0,40.0)(37.0,40.0){3.0} 
\Text(47.0,44.0)[b]{$t$}
\ArrowLine(58.0,40.0)(37.0,40.0) 
\Text(80.0,50.0)[l]{$W^-$}
\DashArrowLine(79.0,50.0)(58.0,40.0){3.0} 
\Text(80.0,30.0)[l]{$\bar{b}$}
\ArrowLine(79.0,30.0)(58.0,40.0) 
\Text(47,0)[b] {diagr.2}
\end{picture} \ 
\begin{picture}(95,79)(0,0)
\Text(15.0,70.0)[r]{$\gamma$}
\DashLine(16.0,70.0)(58.0,70.0){3.0} 
\Text(80.0,70.0)[l]{$t$}
\ArrowLine(58.0,70.0)(79.0,70.0) 
\Text(54.0,60.0)[r]{$t$}
\ArrowLine(58.0,50.0)(58.0,70.0) 
\Text(80.0,50.0)[l]{$\bar{b}$}
\ArrowLine(79.0,50.0)(58.0,50.0) 
\Text(54.0,40.0)[r]{$W^+$}
\DashArrowLine(58.0,30.0)(58.0,50.0){3.0} 
\Text(15.0,30.0)[r]{$\gamma$}
\DashLine(16.0,30.0)(58.0,30.0){3.0} 
\Text(80.0,30.0)[l]{$W^-$}
\DashArrowLine(79.0,30.0)(58.0,30.0){3.0} 
\Text(47,0)[b] {diagr.3}
\end{picture} \ 
\begin{picture}(95,79)(0,0)
\Text(15.0,70.0)[r]{$\gamma$}
\DashLine(16.0,70.0)(58.0,70.0){3.0} 
\Text(80.0,70.0)[l]{$t$}
\ArrowLine(58.0,70.0)(79.0,70.0) 
\Text(54.0,60.0)[r]{$t$}
\ArrowLine(58.0,50.0)(58.0,70.0) 
\Text(80.0,50.0)[l]{$W^-$}
\DashArrowLine(79.0,50.0)(58.0,50.0){3.0} 
\Text(54.0,40.0)[r]{$b$}
\ArrowLine(58.0,30.0)(58.0,50.0) 
\Text(15.0,30.0)[r]{$\gamma$}
\DashLine(16.0,30.0)(58.0,30.0){3.0} 
\Text(80.0,30.0)[l]{$\bar{b}$}
\ArrowLine(79.0,30.0)(58.0,30.0) 
\Text(47,0)[b] {diagr.4}
\end{picture} \ 
\begin{picture}(95,79)(0,0)
\Text(15.0,60.0)[r]{$\gamma$}
\DashLine(16.0,60.0)(37.0,60.0){3.0} 
\Line(37.0,60.0)(58.0,60.0) 
\Text(80.0,70.0)[l]{$\bar{b}$}
\ArrowLine(79.0,70.0)(58.0,60.0) 
\Text(33.0,50.0)[r]{$b$}
\ArrowLine(37.0,60.0)(37.0,40.0) 
\Text(15.0,40.0)[r]{$\gamma$}
\DashLine(16.0,40.0)(37.0,40.0){3.0} 
\Text(47.0,44.0)[b]{$b$}
\ArrowLine(37.0,40.0)(58.0,40.0) 
\Text(80.0,50.0)[l]{$W^-$}
\DashArrowLine(79.0,50.0)(58.0,40.0){3.0} 
\Text(80.0,30.0)[l]{$t$}
\ArrowLine(58.0,40.0)(79.0,30.0) 
\Text(47,0)[b] {diagr.5}
\end{picture} \ 
\begin{picture}(95,79)(0,0)
\Text(15.0,70.0)[r]{$\gamma$}
\DashLine(16.0,70.0)(58.0,70.0){3.0} 
\Text(80.0,70.0)[l]{$\bar{b}$}
\ArrowLine(79.0,70.0)(58.0,70.0) 
\Text(54.0,60.0)[r]{$b$}
\ArrowLine(58.0,70.0)(58.0,50.0) 
\Text(80.0,50.0)[l]{$t$}
\ArrowLine(58.0,50.0)(79.0,50.0) 
\Text(54.0,40.0)[r]{$W^+$}
\DashArrowLine(58.0,30.0)(58.0,50.0){3.0} 
\Text(15.0,30.0)[r]{$\gamma$}
\DashLine(16.0,30.0)(58.0,30.0){3.0} 
\Text(80.0,30.0)[l]{$W^-$}
\DashArrowLine(79.0,30.0)(58.0,30.0){3.0} 
\Text(47,0)[b] {diagr.6}
\end{picture} \ 
\begin{picture}(95,79)(0,0)
\Text(15.0,70.0)[r]{$\gamma$}
\DashLine(16.0,70.0)(58.0,70.0){3.0} 
\Text(80.0,70.0)[l]{$\bar{b}$}
\ArrowLine(79.0,70.0)(58.0,70.0) 
\Text(54.0,60.0)[r]{$b$}
\ArrowLine(58.0,70.0)(58.0,50.0) 
\Text(80.0,50.0)[l]{$W^-$}
\DashArrowLine(79.0,50.0)(58.0,50.0){3.0} 
\Text(54.0,40.0)[r]{$t$}
\ArrowLine(58.0,50.0)(58.0,30.0) 
\Text(15.0,30.0)[r]{$\gamma$}
\DashLine(16.0,30.0)(58.0,30.0){3.0} 
\Text(80.0,30.0)[l]{$t$}
\ArrowLine(58.0,30.0)(79.0,30.0) 
\Text(47,0)[b] {diagr.7}
\end{picture} \ 
\begin{picture}(95,79)(0,0)
\Text(15.0,60.0)[r]{$\gamma$}
\DashLine(16.0,60.0)(37.0,60.0){3.0} 
\Text(47.0,64.0)[b]{$b$}
\ArrowLine(37.0,60.0)(58.0,60.0) 
\Text(80.0,70.0)[l]{$W^-$}
\DashArrowLine(79.0,70.0)(58.0,60.0){3.0} 
\Text(80.0,50.0)[l]{$t$}
\ArrowLine(58.0,60.0)(79.0,50.0) 
\Text(33.0,50.0)[r]{$b$}
\ArrowLine(37.0,40.0)(37.0,60.0) 
\Text(15.0,40.0)[r]{$\gamma$}
\DashLine(16.0,40.0)(37.0,40.0){3.0} 
\Line(37.0,40.0)(58.0,40.0) 
\Text(80.0,30.0)[l]{$\bar{b}$}
\ArrowLine(79.0,30.0)(58.0,40.0) 
\Text(47,0)[b] {diagr.8}
\end{picture} \ 
\begin{picture}(95,79)(0,0)
\Text(15.0,70.0)[r]{$\gamma$}
\DashLine(16.0,70.0)(58.0,70.0){3.0} 
\Text(80.0,70.0)[l]{$W^-$}
\DashArrowLine(79.0,70.0)(58.0,70.0){3.0} 
\Text(54.0,60.0)[r]{$W^+$}
\DashArrowLine(58.0,70.0)(58.0,50.0){3.0} 
\Text(80.0,50.0)[l]{$t$}
\ArrowLine(58.0,50.0)(79.0,50.0) 
\Text(54.0,40.0)[r]{$b$}
\ArrowLine(58.0,30.0)(58.0,50.0) 
\Text(15.0,30.0)[r]{$\gamma$}
\DashLine(16.0,30.0)(58.0,30.0){3.0} 
\Text(80.0,30.0)[l]{$\bar{b}$}
\ArrowLine(79.0,30.0)(58.0,30.0) 
\Text(47,0)[b] {diagr.9}
\end{picture} \ 
\begin{picture}(95,79)(0,0)
\Text(15.0,70.0)[r]{$\gamma$}
\DashLine(16.0,70.0)(58.0,70.0){3.0} 
\Text(80.0,70.0)[l]{$W^-$}
\DashArrowLine(79.0,70.0)(58.0,70.0){3.0} 
\Text(54.0,60.0)[r]{$W^+$}
\DashArrowLine(58.0,70.0)(58.0,50.0){3.0} 
\Text(80.0,50.0)[l]{$\bar{b}$}
\ArrowLine(79.0,50.0)(58.0,50.0) 
\Text(54.0,40.0)[r]{$t$}
\ArrowLine(58.0,50.0)(58.0,30.0) 
\Text(15.0,30.0)[r]{$\gamma$}
\DashLine(16.0,30.0)(58.0,30.0){3.0} 
\Text(80.0,30.0)[l]{$t$}
\ArrowLine(58.0,30.0)(79.0,30.0) 
\Text(47,0)[b] {diagr.10}
\end{picture} \ 
\begin{picture}(95,79)(0,0)
\Text(15.0,60.0)[r]{$\gamma$}
\DashLine(16.0,60.0)(37.0,60.0){3.0} 
\DashLine(37.0,60.0)(58.0,60.0){3.0} 
\Text(80.0,70.0)[l]{$W^-$}
\DashArrowLine(79.0,70.0)(58.0,60.0){3.0} 
\Text(33.0,50.0)[r]{$W^+$}
\DashArrowLine(37.0,60.0)(37.0,40.0){3.0} 
\Text(15.0,40.0)[r]{$\gamma$}
\DashLine(16.0,40.0)(37.0,40.0){3.0} 
\Text(47.0,44.0)[b]{$W^+$}
\DashArrowLine(37.0,40.0)(58.0,40.0){3.0} 
\Text(80.0,50.0)[l]{$\bar{b}$}
\ArrowLine(79.0,50.0)(58.0,40.0) 
\Text(80.0,30.0)[l]{$t$}
\ArrowLine(58.0,40.0)(79.0,30.0) 
\Text(47,0)[b] {diagr.11}
\end{picture} \ 
\begin{picture}(95,79)(0,0)
\Text(15.0,60.0)[r]{$\gamma$}
\DashLine(16.0,60.0)(37.0,60.0){3.0} 
\Text(47.0,64.0)[b]{$W^+$}
\DashArrowLine(37.0,60.0)(58.0,60.0){3.0} 
\Text(80.0,70.0)[l]{$\bar{b}$}
\ArrowLine(79.0,70.0)(58.0,60.0) 
\Text(80.0,50.0)[l]{$t$}
\ArrowLine(58.0,60.0)(79.0,50.0) 
\Text(33.0,50.0)[r]{$W^+$}
\DashArrowLine(37.0,40.0)(37.0,60.0){3.0} 
\Text(15.0,40.0)[r]{$\gamma$}
\DashLine(16.0,40.0)(37.0,40.0){3.0} 
\DashLine(37.0,40.0)(58.0,40.0){3.0} 
\Text(80.0,30.0)[l]{$W^-$}
\DashArrowLine(79.0,30.0)(58.0,40.0){3.0} 
\Text(47,0)[b] {diagr.12}
\end{picture} \ 
\begin{picture}(95,79)(0,0)
\Text(15.0,70.0)[r]{$\gamma$}
\DashLine(16.0,70.0)(37.0,60.0){3.0} 
\Text(15.0,50.0)[r]{$\gamma$}
\DashLine(16.0,50.0)(37.0,60.0){3.0} 
\DashLine(37.0,60.0)(58.0,60.0){3.0} 
\Text(80.0,70.0)[l]{$W^-$}
\DashArrowLine(79.0,70.0)(58.0,60.0){3.0} 
\DashLine(37.0,60.0)(37.0,40.0){3.0} 
\Text(47.0,44.0)[b]{$W^+$}
\DashArrowLine(37.0,40.0)(58.0,40.0){3.0} 
\Text(80.0,50.0)[l]{$\bar{b}$}
\ArrowLine(79.0,50.0)(58.0,40.0) 
\Text(80.0,30.0)[l]{$t$}
\ArrowLine(58.0,40.0)(79.0,30.0) 
\Text(47,0)[b] {diagr.13}
\end{picture} \ 
}
\caption{Diagrams for the top quark production in $\gamma\gamma$ collisions}
\label{aatopt}
\end{figure}
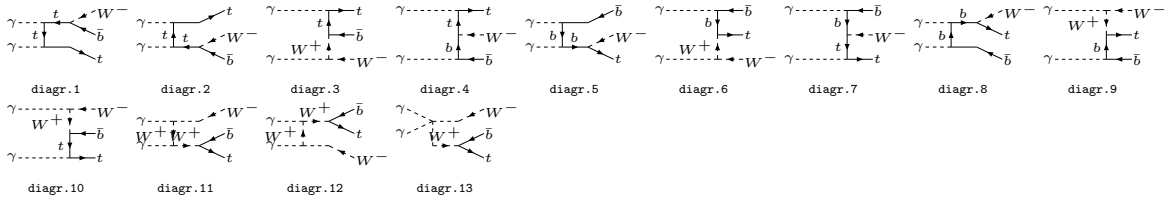

In case of $\gamma\gamma$ collisions there are no nontrivial
gauge invariant subsets Fig.~\ref{aatopt} Ref.\cite{boos-ohl}. A situation
is similar to single top at the LHC in $Wt$ mode. The top pair and single top contributions could be separated
in a gauge invariant way as above.

In $\gamma e$ collisions the top quarks could be produced
only singly Fig.~\ref{aetopt}, the corresponding diagrams are shown in the following
figure.
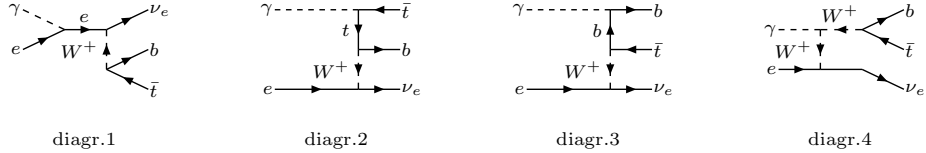
\begin{figure}[!h!]
{
\unitlength=0.75 pt
\SetScale{0.75}
\SetWidth{0.7}      
\scriptsize    
{} \qquad\allowbreak
\begin{picture}(95,79)(0,0)
\Text(15.0,70.0)[r]{$\gamma$}
\DashLine(16.0,70.0)(37.0,60.0){3.0} 
\Text(15.0,50.0)[r]{$e$}
\ArrowLine(16.0,50.0)(37.0,60.0) 
\Text(47.0,64.0)[b]{$e$}
\ArrowLine(37.0,60.0)(58.0,60.0) 
\Text(80.0,70.0)[l]{$\nu_e$}
\ArrowLine(58.0,60.0)(79.0,70.0) 
\Text(54.0,50.0)[r]{$W^+$}
\DashArrowLine(58.0,40.0)(58.0,60.0){3.0} 
\Text(80.0,50.0)[l]{$b$}
\ArrowLine(58.0,40.0)(79.0,50.0) 
\Text(80.0,30.0)[l]{$\bar{t}$}
\ArrowLine(79.0,30.0)(58.0,40.0) 
\Text(47,0)[b] {diagr.1}
\end{picture} \ 
{} \qquad\allowbreak
\begin{picture}(95,79)(0,0)
\Text(15.0,70.0)[r]{$\gamma$}
\DashLine(16.0,70.0)(58.0,70.0){3.0} 
\Text(80.0,70.0)[l]{$\bar{t}$}
\ArrowLine(79.0,70.0)(58.0,70.0) 
\Text(54.0,60.0)[r]{$t$}
\ArrowLine(58.0,70.0)(58.0,50.0) 
\Text(80.0,50.0)[l]{$b$}
\ArrowLine(58.0,50.0)(79.0,50.0) 
\Text(54.0,40.0)[r]{$W^+$}
\DashArrowLine(58.0,50.0)(58.0,30.0){3.0} 
\Text(15.0,30.0)[r]{$e$}
\ArrowLine(16.0,30.0)(58.0,30.0) 
\Text(80.0,30.0)[l]{$\nu_e$}
\ArrowLine(58.0,30.0)(79.0,30.0) 
\Text(47,0)[b] {diagr.2}
\end{picture} \ 
{} \qquad\allowbreak
\begin{picture}(95,79)(0,0)
\Text(15.0,70.0)[r]{$\gamma$}
\DashLine(16.0,70.0)(58.0,70.0){3.0} 
\Text(80.0,70.0)[l]{$b$}
\ArrowLine(58.0,70.0)(79.0,70.0) 
\Text(54.0,60.0)[r]{$b$}
\ArrowLine(58.0,50.0)(58.0,70.0) 
\Text(80.0,50.0)[l]{$\bar{t}$}
\ArrowLine(79.0,50.0)(58.0,50.0) 
\Text(54.0,40.0)[r]{$W^+$}
\DashArrowLine(58.0,50.0)(58.0,30.0){3.0} 
\Text(15.0,30.0)[r]{$e$}
\ArrowLine(16.0,30.0)(58.0,30.0) 
\Text(80.0,30.0)[l]{$\nu_e$}
\ArrowLine(58.0,30.0)(79.0,30.0) 
\Text(47,0)[b] {diagr.3}
\end{picture} \ 
{} \qquad\allowbreak
\begin{picture}(95,79)(0,0)
\Text(15.0,60.0)[r]{$\gamma$}
\DashLine(16.0,60.0)(37.0,60.0){3.0} 
\Text(47.0,64.0)[b]{$W^+$}
\DashArrowLine(58.0,60.0)(37.0,60.0){3.0} 
\Text(80.0,70.0)[l]{$b$}
\ArrowLine(58.0,60.0)(79.0,70.0) 
\Text(80.0,50.0)[l]{$\bar{t}$}
\ArrowLine(79.0,50.0)(58.0,60.0) 
\Text(33.0,50.0)[r]{$W^+$}
\DashArrowLine(37.0,60.0)(37.0,40.0){3.0} 
\Text(15.0,40.0)[r]{$e$}
\ArrowLine(16.0,40.0)(37.0,40.0) 
\Line(37.0,40.0)(58.0,40.0) 
\Text(80.0,30.0)[l]{$\nu_e$}
\ArrowLine(58.0,40.0)(79.0,30.0) 
\Text(47,0)[b] {diagr.4}
\end{picture} \ 
}
\caption{Diagrams for the top quark production in $\gamma e$ collisions}
\label{aetopt}
\end{figure}
This is one of so called "gold plated" processes in $\gamma e$
collision mode of ILC.

Table~\ref{tab:lc_cs} shows single top production cross sections in various collision modes for unpolarized and polarized beams expected at a linear 
collider\cite{Boos:2001sj}.

\begin{table}[!h!]
\begin{center}
\tbl{ Top quark production by $e^+$, $e^-$ and $\gamma$ beams with
various polarizations. The single top quark production cross sections 
(with $m_{top}=$175 GeV) are
given for the channels
 $e^- e^+, \gamma \gamma \to e^- \bar \nu_e t \bar b$,
 $e^- e^- \to e^- \nu_e \bar t b$ and $\gamma e^- \to \nu_e \bar t b$
at $\sqrt{s}=$0.5 and 1.0 TeV.}
{\begin{tabular}{|c|c|c|c|c|c|}
\hline
 beams & No. of& polarization & $t \bar t$ & $\sigma_{single \;
top}$, fb &
$\sigma_{single \; top}$, fb \\
       & diagrams & and subset & production &  $\sqrt{s}=$0.5 TeV &
                                     $\sqrt{s}=$1 TeV \\
\hline
 $e^- e^+$ & 20 & unpol       &  yes &  3.1 & 6.7\\
           & 10 & unpol,$s$-ch.& yes & 2.3 & 2.3 \\
           & 10 & unpol,$t$-ch.& no                &  0.8 & 4.4 \\
           & 20 & LR          &  yes & 10.0& 16.9\\
           & 11 & RL          &  yes & 1.7 & 1.0 \\
           &  9 & RR          &  no             &  1.0 & 8.1\\
           &  2 & LL          &  no                & - & - \\
\hline
 $e^- e^-$ & 20 & unpol       &  no              & 1.7 & 9.1 \\
           & 20 & LL          &  no              & 2.6 & 19.1 \\
           & 11 & LR          &  no                & 2.1 & 14.0 \\
           & 11 & RL          &  no                & 2.1 & 14.0 \\
           &  4 & RR          &  no                & 0.02 & 0.96 \\
\hline
$\gamma e^-$ &4 & unpol       &  no                & 30.3 & 67.6 \\
             &4 & $-$L    & no                &  38.9 & 121.3 \\
             &4 & +L    & no                &  94.3 & 174.7 \\
\hline
$\gamma \gamma$& 21& unpol    & yes  & 9.2 & 18.8\\
           & 21 &(++)         & yes  & 11.1 & 19.2\\
           & 21 &($-$ $-$)     & yes  & 7.9 & 15.7\\
           & 21 &(+$-$) or ($-$+) & yes & 8.5& 19.2\\
\hline
       
\end{tabular}\label{tab:lc_cs}}
\end{center}
\end{table}
One can see from the Table 1 that both t-channel and s-channel parts contribute to the single top production rate. 
However the t-channel contribution grows with energy and dominates for very high energies. The 
 NLO QCD corrections to the single top quark production in an effective $W\gamma$ approximation (EWA) have been computed in Ref~\refcite{Kuhn:2003pn} in the case of $\gamma e$ and in Ref.~\refcite{Penunuri:2011hp} for the case of $e^+ e^-$ collisions showing the corrections are of the order of $10\%$.

\begin{figure}[!h!]
\centerline{\psfig{file=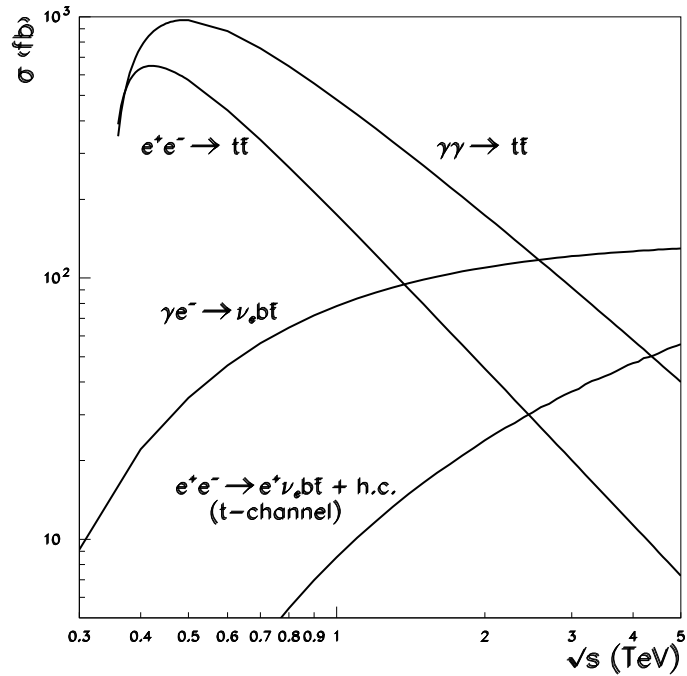,width=10cm}}
\caption{Cross sections for the top quark production in various processes as a function of linear collider energies\label{fig:LC_xsections}}
\end{figure}

Cross sections of various processes in different collision modes at a linear collider  are collected in Fig.~\ref{fig:LC_xsections}
\cite{Weiglein:2004hn}.
One can stress the single top quark production in $\gamma e$ collisions
is smaller than the top-pair rate in $e^+e^-$
only by a factor of 1/8 at 500--800 GeV energies, and it becomes the
dominant LC process for the top production at a multi-TeV LC like CLIC. 
The $|V_{tb}|$ matrix element can be measured at a LC, especially in $\gamma e$ collisions,
significantly more accurate compared to the LHC.

\begin{figure}[!h!]
\begin{center}   
\centerline{\psfig{file=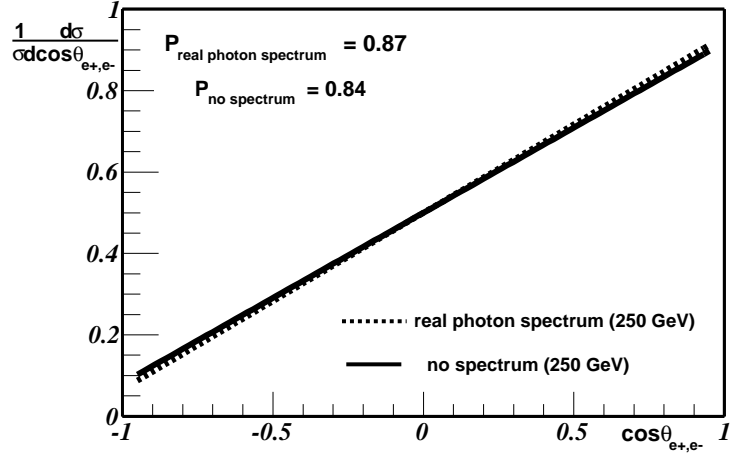,width=10cm}}
\caption{Angular distribution for the angle between the
electron beam and muon from the top decay in single-antitop-quark
production at $\gamma e^-$ 500 GeV collider.}
\label{fig:spincorr_LC.eps}
\end{center}
\end{figure}

Similar to single top production in $tW$ more at hadron colliders there is interesting spin correlation 
between top production and decay in $\gamma e$ collisions. 
 Here the directions of the initial photon
and the electrons play the role of the gluon and lepton direction in
the $tW$ process at a hadron collider. The directions of $\gamma$ and
electron beams are close to the top-quark rest frame since the top is
moving slowly here. So one would expect that the top quark is strongly
polarized in the direction of the initial electron beam. Indeed, the
angular distribution for the angle between the lepton from the top
decay and the initial electron beam shows about 90\% correlation
\cite{Boos:2002xw} (see Fig.\ \ref{fig:spincorr_LC.eps}). 

\section{``New Physics'' via single top. }

Single top processes having a significant production rate at both the Tevatron and the LHC colliders are important backgrounds for various searches beyond the Standard Model (BSM). However, the processes are extremely interesting for different types of possible manifestations of ``New Physics''. In general, two main situations are possible depending on relations between characteristic collision energy and thresholds of possible new states. If the collision energy is smaller than the production threshold $E_{\rm collisions} < E_{threshold}$ new states can not be produced directly and ``New Physics'' may manifest as deviations in production cross sections and kinematic distributions due to possible anomalous couplings or the interference of new resonances with the SM contribution below thresholds. In case of single top, the anomalous couplings could be anomalous $Wtb$ and/or FCNC couplings. If the collision energy is greater than the production threshold $E_{\rm collisions} > E_{threshold}$ new states can be 
produced directly in the production or in the decay of the top quark. Such states could be an additional vector ($W^\prime$) or scalar ($H^+$)  bosons predicted in many extensions of SM.

Generic parametrisation of the anomalous couplings follows from the effective Lagrangian approach\cite{Buchmuller:1985jz}.
 There are number of dimension 6 effective operators which preserve the SM gauge invariance\cite{AguilarSaavedra:2008zc,Zhang:2012cd} and lead to modifications of $Wtb$ vertex and appearance of additional FCNC couplings of the top quark with $u-$ and $c$-quarks.   

The effective $Wtb$ vertex has the following form\cite{Kane:1991bg}:     
\begin{eqnarray}
 \mathcal{L}&=& \frac{g}{\sqrt{2}}\bar{b} \gamma^\mu V_{tb}
\nonumber (f^L_V P_L + f^R_V P_R) t W_{\mu}^{-}\\
& &-\frac{g}{\sqrt{2}} \bar{b} \frac{i\sigma^{\mu\nu} q_{\nu}V_{tb}}{M_W} 
(f^L_T P_L + f^R_T P_R) t W_{\mu}^{-} + h.c. \, ,
\label{anom_wtb_eq_lagrangian}
\end{eqnarray}
 where  $M_W$ is the mass of the $W$~boson, $q$ is its four-momentum, $V_{tb}$ is the 
Cabibbo-Kobayashi-Maskawa matrix element and $P_{L}=(1 - \gamma_5)/2$ 
($P_{R}=(1 + \gamma_5)/2$) is the left-handed (right-handed) projection operator. The anomalous couplings  $f^L_V,f^R_V,f^L_T,f^R_T$ are related to the constants in front of the effective operators\cite{AguilarSaavedra:2008zc} in the following way:
\begin{eqnarray}
\label{eq:operator}
|f^L_V| &=& 1 + |C_{\phi q}^{(3,3+3)}| \frac{v^2}{V_{tb} \Lambda^2} \; ,\nonumber \\
|f^R_V| &=& \frac{1}{2} |C_{\phi \phi}^{33}| \frac{v^2}{V_{tb} \Lambda^2} \; ,\nonumber \\
|f^L_T| &=& \sqrt{2} |C_{dW}^{33}| \frac{v^2}{V_{tb} \Lambda^2} \; ,\nonumber \\
|f^R_T| &=& \sqrt{2} |C_{uW}^{33}| \frac{v^2}{V_{tb} \Lambda^2} \; , 
\end{eqnarray}
 One should mentioned that the couplings $C$ are naturally of the order of unity. Therefore, one may expect the natural size of $f$ couplings is of the order of $\frac{v^2}{\Lambda^2}$ which is about 0.05 or less. The theoretical estimations performed in Refs.\cite{Boos:1999dd,Tsuno:2005qb,AguilarSaavedra:2008gt,Bernreuther:2008us} have shown the Tevatron and LHC collider potentials showing the natural size of the parameters could be achieved. However, the expected limits could not be much better because uncertainties are dominated by systematics. Expected bounds could be improved by a factor of $2 \div 3$ at a Linear Collider specially if the $e\gamma$-collision mode will be realised\cite{Boos:2001sj,Boos:1997rd}. One should mentioned that in the single top processes the $Wtb$ anomalous couplings contribute to the production, to the subsequent decay of produced top quark and change the total width of the top quark correspondingly. The resulting dependence of the signal process on anomalous parameters is more 
complicated than a simple polinom structure. Such a dependence affects spin correlations between production and decay and can be exploited in experimental analysis. In the pair top quark production processes, anomalous operators are detectable in the W boson helicity fractions in the decay of top quarks\cite{Chen:2005vr}.
 The recent and most tight direct limits\footnote{Indirect limits on the anomalous $Wtb$ couplings follow from measurements of the $b \rightarrow s\gamma$ decays\cite{Drobnak:2011aa}, however, the limits are obtained with the assumption that there are no other BSM contribution in the loop.} to the anomalous couplings in $Wtb$ vertex comes from combination of the measurements in single and pair top quark production processes\cite{:2012iwa}. 
 The current limits are\cite{:2012iwa}: 
$|f_V^R|^2 < 0.30 $, 
$|f_T^L|^2 < 0.05 $,
$|f_T^R|^2 < 0.12 $.
Assuming the scale $\Lambda=1$~TeV these limits translate to the corresponding limits on anomalous operators to be $|C_{\phi q}^{(3,3+3)}| < 14.7$, $|C_{\phi \phi}^{33}|< 18.0$, $|C_{dW}^{33}|< 2.5$, and $|C_{uW}^{33}|< 4.1$.
 
\begin{figure}[!h!tbp]
\vspace{-0.07in}
  \begin{center} 
 \includegraphics[width=0.499\textwidth]{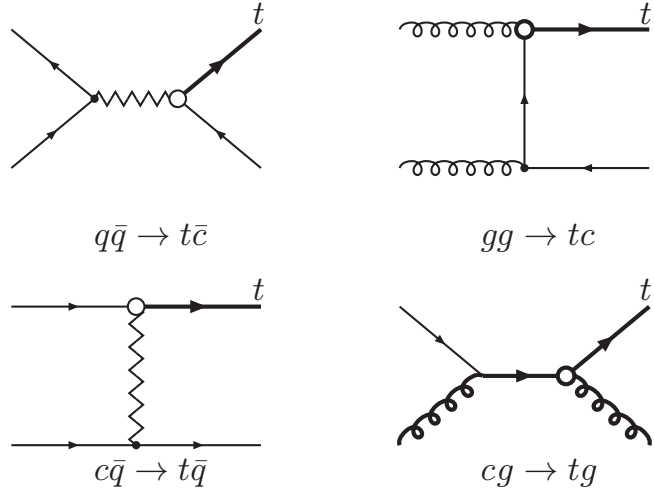}
  \end{center} 
\vspace{-0.1in}
\caption{The representative leading order Feynman diagrams for FCNC
gluon coupling between a charm quark and a top quark (for a u-quark the diagrams are the same).
\label{fig:fcnc_feynman} }
\end{figure}

    FCNC top quark anomalous couplings can be probed in their production or in their rear decays\cite{Beneke:2000hk}. In particular, $tug$ and $tcg$ FCNC couplings may affect the single top production rate and it was exploited at the Tevatron and the LHC to set the corresponding limits. The gauge invariant effective Lagrangian describing these anomalous couplings has the following form 
\begin{equation}
\label{eq:fcnc_lagrangian}
{\cal L}_{\rm FCNC} = \frac{\kappa_{tgq}}{\Lambda} g_s \bar{q} \sigma^{\mu\nu} \frac{\lambda^a}{2} t G^a_{\mu\nu} ,
\end{equation}
where $q$ = $u$ or $c$, with $u$, $c$ and $t$ representing the quark fields; $\kappa_{tgq}$ defines the strength of the $tgu$ or $tgc$ couplings; $g_s$ and $\lambda^a$ are the strong coupling constant and color matrices; $\sigma^{\mu\nu}=i/2(\gamma_{\mu}\gamma_{\nu}-\gamma_{\nu}\gamma_{\mu})$ and $G^a_{\mu\nu}$ are the Dirac tensor and the gauge field tensor of the gluon.
From the above effective Lagrangian one easily obtains the partial rare decay widths $t\to qg$:
\begin{eqnarray} 
\Gamma({ t \to q g}) \,\,  &=& \left( \frac{\kappa^g_{tq}}{\Lambda} \right)^2 
  \frac{8}{3} \alpha_s m_t^3  \quad \quad , \quad \quad 
\label{anomeq:fcnc_br} 
\end{eqnarray} 
The partial width is directly proportional to the ratio $\left( \frac{\kappa^g_{tq}}{\Lambda} \right)^2 $, therefore various authors and collaborations present limits either in terms of the constants $\kappa$ or in terms of corresponding branching ratio.
The representative diagrams contributing to various processes\cite{Han:1998tp,Liu:2005dp,Gao:2009rf} of single top production due to presence of FCNC interactions are shown in Fig.~\ref{fig:fcnc_feynman}.
The model independent analysis based on the signal diagrams in Fig.~\ref{fig:fcnc_feynman} corresponds to the case when extra hard jet is detected in association with the top quark. The most recent results on FCNC anomalous coupling limits in terms of both couplings and branchings  are presented in Table~\ref{tab:obslim}. The accumulated luminosities for each of the analysis are listed explicitly in the table.  From the table one can see that the current LHC limits at 7 TeV obtained by the ATLAS collaboration are already significantly tighter than the Tevatron limits by D0 and CDF experiments with almost the same luminosity. One should stress the best current limits are far above very small SM value for the decay branching ratio $Br^{SM}(t\to cg) \approx 5 \times 10^{-11}$, however, being in the interesting range for some extensions of the SM\cite{Beneke:2000hk}. 
\begin{table}[!h!t]
\tbl{ Observed 95\% C.L. upper limits on anomalous FCNC couplings and corresponding branching ratios.}
{\begin{tabular}{@{}cccc@{}} \toprule
Parameter  & D0   (1.96 TeV, 2.3~fb$^{-1}$)\cite{Abazov:2010qk}  &  CDF    (1.96 TeV, 2.2~fb$^{-1}$)\cite{Aaltonen:2008qr}  & ATLAS (7 TeV, 2.05~fb$^{-1}$)\cite{Aad:2012gd}\\ \hline
$\kappa^g_{tu} / \Lambda $ & $0.013$~TeV$^{-1}$ & & $6.9\times10^{-3}$~TeV$^{-1}$ \\ \hline
$\kappa^g_{tc} / \Lambda $ & $0.057$~TeV$^{-1}$ & & $1.6\times10^{-2}$~TeV$^{-1}$ \\ \hline
$Br(t\to ug) $ & $2.0\times10^{-4}$ & $3.9\times10^{-4}$ & $5.7\times10^{-5}$ \\ \hline
$Br(t\to cg) $ & $3.9\times10^{-3}$ & $5.7\times10^{-3}$ & $2.7\times10^{-4}$ \\ \hline
\hline
\end{tabular}\label{tab:obslim}}
\end{table}

As was mentioned, if the achievable characteristic collision energy $E_{\rm collisions}$  is grater than a production threshold of possible new particles $E_{\rm threshold}$ one may detect these particles directly. 
In our case of single top production, such particles  could be either newly charged vector or scalar bosons produced as $s-$channel resonances decaying to the top quark. We consider only single top production processes and, therefore, we refer only to $W^\prime$ and charged scalar decaying to top and bottom quarks and do not discuss the processes with leptonic decays  of $W^\prime$ where the observed limits are much stronger in some cases. 
Such models like Non-Commuting Extended Technicolor
\cite{Chivukula:1995gu},
Composite\cite{Kaplan:1983fs,Georgi:1984af} and Little higgs
models\cite{Arkani:2001nc,Kaplan:2003uc,Schmaltz:2005ky}, models of composite gauge
bosons\cite{composites}, Supersymmetric top-flavor models\cite{Batra:2003nj}, Grand Unification\cite{guts},
Superstring theories\cite{Cvetic:1996mf,Pati:2006nw,superstrings}, and Left-Right symmetric
models\cite{Pati:1974yy}-\cite{Langacker:1989xa},
 represent examples where extension of gauge group lead to
appearing of $W^{\prime}$. $W^{\prime}$ appears also in the  Universal extra-dimension\cite{Datta:2000gm,extradimensions} type of models. Charged scalar (pseudo-scalar) bosons naturally present in many SM extensions as well, in particular, in two-Higgs-doublet models (2HDMs)  of various  types two charged Higgs bosons appear. 

In searches for these new bosons in a model independent way, the effective Lagrangian approach is used. Corresponding effective Lagrangians have the following forms~\ref{eq:wprime_lagrangian} and ~\ref{eq:higgs_lagrangian} neglecting higher-dimensional structures.

\begin{eqnarray}
\label{eq:wprime_lagrangian}
{\cal L} = \frac{V_{q_iq_j}}{2\sqrt{2}} g_w \overline{q}_i\gamma_\mu 
\bigl( a^R_{q_iq_j} (1+{\gamma}^5) + a^L_{q_iq_j}
(1-{\gamma}^5)\bigr) W^{\prime\mu} q_j 
+ \mathrm{H.c.} \,, 
\end{eqnarray}
where $a^R_{q_iq_j}, a^L_{q_iq_j}$ - left and right couplings of $W^\prime$ 
to quarks,
$g_w = e/(s_w)$ is the SM weak coupling constant and 
$V_{q_iq_j}$ is the SM CKM matrix element.
The notations are taken such that for so-called 
SM-like $W^{\prime}$ $a^L_{q_iq_j}=1$ and 
$a^R_{q_iq_j}=0$.

\begin{eqnarray}
\label{eq:higgs_lagrangian}
\mathcal{L} & = & \frac{g_w V_{q_i q_j}}{2\sqrt{2}} H^+ \bar{q}_i \left[
  g^{ij}_L \left( \frac{1 - \gamma^5}{2} \right) + g^{ij}_R \left( \frac{1 + \gamma^5}{2}
  \right) \right] q_j 
\end{eqnarray}
Since there are no charge scalars in SM the couplings $g_L^{ij}$, $g_R^{ij}$ are obviously equal to zero in SM.

  For the $W^\prime$ searches the limits on $W^\prime$ mass depend on both left and right couplings $a^L, a^R$ and for the case of right-handed interaction of $W^\prime$ on a relation between the $W^\prime$ mass and the mass of possible right-handed neutrino $\nu_R$. If the decay of $W^\prime$ to right-handed neutrino is kinematically allowed the branching ratio of $W^\prime$ decay to top and bottom quarks is smaller and corresponding mass limits are expected to be worse. The NLO corrections to the $s-$channel $W^\prime$ production with subsequent decay were computed in Ref.\cite{Sullivan:2002jt} and an influence of  $W^\prime$--$W^{SM}$ interference was demonstrated in Ref.\cite{Boos:2006xe}.
In case the $W^\prime$ is a first KK state in models with extra deminsions, an additional interference with rest of KK tower should be taken into account Ref.\cite{Boos:2011ib}.
 The 95\% C.L. limits obtained by the Tevatron at $\sqrt{s}=1.96$ TeV and the LHC collaborations at  $\sqrt{s}=7$ TeV are collected in the Table~\ref{tab:wprime} for the cases where the couplings $a^L, a^R$ are equal to $1, 0$ (Left), $0, 1$ (Right) and $1, 1$ (Left, Right).
  One can see the ATLAS limit with about two times smaller statistics is already better than the Tevatron limits.

\begin{table}[!h!t]
\tbl{ Observed 95\% C.L. limits on the lowest bound of the mass of $W^\prime$ boson.
}
{\begin{tabular}{l|c|c|c|c}
\hline \hline
Scenario  & D0   (2.3~fb$^{-1}$)\cite{Abazov:2011xs}  &  CDF    (1.9~fb$^{-1}$)\cite{Aaltonen:2009qu} & CMS (5~fb$^{-1}$)\cite{CMS-PAS-EXO-12-001}  & ATLAS (1.04~fb$^{-1}$)\cite{Aad:2012ej}\\ \hline
$M_{W^\prime}{\rm (Left)}$     & 863 GeV &        &          &  \\ \hline
$M_{W^\prime}{\rm (Right)}, M_{W^\prime}>M_{\nu_R}$     & 885 GeV & 800 GeV& 1.85 TeV &  \\ \hline
$M_{M^\prime}{\rm (Right)}, M_{W^\prime}<M_{\nu_R}$           & 890 GeV & 825 GeV&          & 1.13 TeV\\ \hline
$M_{W^\prime}{\rm (Left,Right)}, M_{W^\prime}>M_{\nu_R}$ & 916 GeV &        &          &  \\ \hline
\hline
\end{tabular}\label{tab:wprime}}
\end{table}

The available experimental limits for the charged scalar Higgs particle decaying to top and bottom quarks are published by D0 collaboration\cite{Abazov:2008rn}. The mass limits are given in the mass range grater than 180 GeV ($M_{H^+}>M_{top} + M_{b}$) since lower mass range is covered with better precision from the top pair production with the top decay to the charged Higgs and b-quark. The limits strongly depend on the ratio of two Higgs vacuum expectation values $\tan\beta$ and the type of  2HDMs (Type I, Type II and Type III).

\section{Conclusions and outlook}

The top quark, being the heaviest elementary particle discovered so far, has been observed in pair first and later in single production mode. Single top quark registration at the Tevatron and then at the LHC is an important confirmation of the structure of the electroweak part of the Standard Model. Measured production cross sections in combination of $t-$ and $s-$channel modes at the Tevatron and in the $t-$channel only at the Tevatron and the LHC  are in a good agreement with the SM computations at NLO level including part of NNLO corrections. $tW-$channel is very small to be measured at the Tevatron while it was measured for the first time at the LHC. Within rather large experimental uncertainties the results for the $tW-$channel are in an agreement with SM as well. The $s-$channel cross section at the LHC is much smaller than the $t-$channel making an observation of this channel at the LHC very challenging especially because the $t-$channel gives a huge irreducible background to the $s-$channel in a 
kinematic range where an additional light jet is undetectable.  However, one expects the detection of the $s-$channel with increased analyzed statistic. 
Having small rates, these processes are interesting to search for deviations from the SM predictions.

The performed studies of the single top production allowed to measure directly for the first time the CKM mixing matrix parameter $|V_{tb}|$. Published results are in agreement with expected unity, however, still at about 20\% level of accuracy.  

  The top is very heavy and as the result is very special fermion in SM. The single top production is a power tool to search for delicate deviations from the SM. No such deviations have been observed yet and limits on the anomalous $Wtb$ and FCNC top quark couplings as well as on parameters of new vector and scalar bosons decaying to top quark were extracted. 
  The mass limits for new resonances are expected to be much higher (in few TeV range) with increasing LHC energy. While the limits on anomalous couplings will be dominated by systematic uncertainties and therefore might be improved by a factor of two or so. Much tighter limits at a percentage level of accuracy could be achieved at a future Linear Collider. 

\section*{Acknowledgments}
The authors are grateful to many theory colleagues and colleagues from single top groups of D0 and CMS collaborations for clarifying discussions and joined studies. The work was supported in part by the RFBR grant 12-02-93108-CNRSL−a and the grant of Russian Ministry of Education and Science NS-3920.2012.2.


\begin{thebibliography}{000}
\bibitem{top_cdf_1995_ttbar}
F.~Abe et al., [CDF Collaboration] \\
Phys.\ Rev.\ Lett.\ {\bf 74}, 2626 (1995) [arXiv:hep-ex9503002].

\bibitem{top_d0_1995_ttbar}
S.~Abachi et al., [D0 Collaboration] \\
Phys.\ Rev.\ Lett.\ {\bf 74}, 2632 (1995) [arXiv:hep-ex9503003].

\bibitem{Blondel:1996wm} 
  A.~Blondel,
  Nuovo Cim.\ A {\bf 109}, 771 (1996). CERN Report LEPEWWG/96-02.

\bibitem{Abazov:2009ii}
  V.~M.~Abazov {\it et al.} [ D0 Collaboration ],
  Phys.\ Rev.\ Lett.\  {\bf 103}, 092001 (2009).
  [arXiv:0903.0850 [hep-ex]].

\bibitem{Aaltonen:2009jj}
  T.~Aaltonen {\it et al.} [ CDF Collaboration ],
  Phys.\ Rev.\ Lett.\  {\bf 103}, 092002 (2009).
  [arXiv:0903.0885 [hep-ex]].


\bibitem{1007.3178:1900yx}
   [CDF and D0 Collaboration],
  arXiv:1007.3178 [hep-ex].

\bibitem{Jezabek:1993wk} 
  M.~Jezabek and J.~H.~Kuhn,
  Phys.\ Rev.\ D {\bf 48}, 1910 (1993)
  [Erratum-ibid.\ D {\bf 49}, 4970 (1994)]
  [hep-ph/9302295].
  M.~Jezabek and J.~H.~Kuhn,
  Nucl.\ Phys.\ B {\bf 314}, 1 (1989).

  
\bibitem{Bigi:1986jk}
I.~I.~Bigi, Y.~L.~Dokshitzer, V.~A.~Khoze, J.~H.~Kuhn and P.~M.~Zerwas,
Phys.\ Lett.\ B {\bf 181} (1986) 157.

\bibitem{Jezabek:1994zv}
  M.~Jezabek and J.~H.~Kuhn,
  Phys.\ Lett.\  B {\bf 329}, 317 (1994)
  [arXiv:hep-ph/9403366].


\bibitem{Mahlon:1996pn}
  G.~Mahlon and S.~J.~Parke,
  Phys.\ Rev.\ D { 55}, 7249 (1997)
  [hep-ph/9611367]; 
  G.~Mahlon and S.~J.~Parke,
  Phys.\ Lett.\ B { 476}, 323 (2000)
  [hep-ph/9912458].

\bibitem{Boos:2002xw}
  E.~E.~Boos and A.~V.~Sherstnev,
  Phys.\ Lett.\ B { 534}, 97 (2002) 
  [hep-ph/0201271].

  
\bibitem{Beneke:2000hk}
  M.~Beneke {\it et al.},
  ``Top quark physics,''
  arXiv:hep-ph/0003033.

\bibitem{Tait:2000sh}
  T.~M.~P.~Tait and C.~P.~P.~Yuan,
  Phys.\ Rev.\  D {\bf 63}, 014018 (2000)
  [arXiv:hep-ph/0007298].

\bibitem{Willenbrock:2002ta} 
  S.~Willenbrock,
  hep-ph/0211067.
 
\bibitem{Chakraborty:2003iw} 
  D.~Chakraborty, J.~Konigsberg and D.~L.~Rainwater,
  Ann.\ Rev.\ Nucl.\ Part.\ Sci.\  {\bf 53}, 301 (2003)
  [hep-ph/0303092].

\bibitem{Kehoe:2007px} 
  R.~Kehoe, M.~Narain and A.~Kumar,
  Int.\ J.\ Mod.\ Phys.\ A {\bf 23}, 353 (2008)
  [arXiv:0712.2733 [hep-ex]].

 
\bibitem{Heinson:2011jn} 
  A.~Heinson and T.~R.~Junk,
  Ann.\ Rev.\ Nucl.\ Part.\ Sci.\  {\bf 61}, 171 (2011)
  [arXiv:1101.1275 [hep-ex]].
  
\bibitem{Schilling:2012dx} 
  F.~-P.~Schilling,
  Int.\ J.\ Mod.\ Phys.\ A {\bf 27}, 1230016 (2012)
  [arXiv:1206.4484 [hep-ex]].

  
\bibitem{Willenbrock:1986cr}
 S.~S.~D.~Willenbrock and D.~A.~Dicus,
  Phys.\ Rev.\ D { 34}, 155 (1986);

  \bibitem{Yuan:1989tc}
  C.~P.~Yuan,
  Phys.\ Rev.\ D { 41}, 42 (1990);

  \bibitem{Jikia:1991yd}
  G.~V.~Dzhikia and S.~R.~Slabospitsky,
  Sov.\ J.\ Nucl.\ Phys.\  { 55}, 1387 (1992);
  Phys.\ Lett.\ B { 295}, 136 (1992);

  \bibitem{Cortese:1991fw}
  S.~Cortese and R.~Petronzio,
  Phys.\ Lett.\ B { 253}, 494 (1991);

  \bibitem{Ellis:1992yw}
  R.~K.~Ellis and S.~J.~Parke,
  Phys.\ Rev.\ D { 46}, 3785 (1992);

  \bibitem{Bordes:1992sv}
  G.~Bordes and B.~van Eijk,
  Z.\ Phys.\ C { 57}, 81 (1993);

  \bibitem{Carlson:1994bg}
  D.~O.~Carlson, E.~Malkawi, and C.~P.~Yuan,
  Phys.\ Lett.\ B { 337}, 145 (1994)
  [hep-ph/9405277];

  \bibitem{Stelzer:1995mi}
  T.~Stelzer and S.~Willenbrock,
  Phys.\ Lett.\ B { 357}, 125 (1995)
  [hep-ph/9505433];


  \bibitem{Atwood:1996pd}
  D.~Atwood, S.~Bar-Shalom, G.~Eilam, and A.~Soni,
  Phys.\ Rev.\ D { 54}, 5412 (1996)
  [hep-ph/9605345];

  \bibitem{Li:1996ir}
 C.~S.~Li, R.~J.~Oakes and J.~M.~Yang,
 Phys.\ Rev.\ D { 55}, 5780 (1997)
 [hep-ph/9611455];

\bibitem{Heinson:1996zm}
  A.~P.~Heinson, A.~S.~Belyaev, and E.~E.~Boos,
  Phys.\ Rev.\ D { 56}, 3114 (1997)
  [hep-ph/9612424].

  \bibitem{Bar-Shalom:1997sj}
  S.~Bar-Shalom, G.~Eilam, A.~Soni, and J.~Wudka,
  Phys.\ Rev.\ D { 57}, 2957 (1998)
  [hep-ph/9708358];

  \bibitem{Tait:1997fe}
 T.~Tait and C.~P.~Yuan,
  hep-ph/9710372.

\bibitem{Belyaev:1998dn}
  A.~S.~Belyaev, E.~E.~Boos, and L.~V.~Dudko,
  Phys.\ Rev.\ D { 59}, 075001 (1999)
  [hep-ph/9806332].

  \bibitem{Bordes:1994ki}
  G.~Bordes and B.~van Eijk,
  Nucl.\ Phys.\ B { 435}, 23 (1995);
  
   \bibitem{Pittau:1996rp}
  R.~Pittau,
  Phys.\ Lett.\ B { 386}, 397 (1996)
  [hep-ph/9603265];

 
\bibitem{Smith:1996ij}
  M.~C.~Smith and S.~Willenbrock,
  Phys.\ Rev.\ D { 54}, 6696 (1996)
  [hep-ph/9604223].

  \bibitem{Li:1996bh}
  C.~S.~Li, R.~J.~Oakes, and J.~M.~Yang,
  Phys.\ Rev.\ D { 55}, 1672, 5780 (1997)
  [hep-ph/9608460],[hep-ph/9611455];

  \bibitem{Stelzer:1997ns}
  T.~Stelzer, Z.~Sullivan, and S.~Willenbrock,
  Phys.\ Rev.\ D { 56}, 5919 (1997)
  [hep-ph/9705398];

 \bibitem{Stelzer:1998ni}
 T.~Stelzer, Z.~Sullivan, and S.~Willenbrock,
  Phys.\ Rev.\ D { 58}, 094021 (1998)
  [hep-ph/9807340].

\bibitem{Zhu:2001hw} 
  S.~Zhu
  hep-ph/0109269.
  
 \bibitem{Harris:2002md}
  B.~W.~Harris {\it et al.},
  Phys.\ Rev.\ D { 66}, 054024 (2002)
  [hep-ph/0207055].

  \bibitem{Sullivan:2004ie}
  Z.~Sullivan,
  Phys.\ Rev.\ D { 70}, 114012 (2004)
  [hep-ph/0408049].

\bibitem{Campbell:2004ch}
  J.~Campbell, R.~K.~Ellis, and F.~Tramontano,
  Phys.\ Rev.\ D { 70}, 094012 (2004)
  [hep-ph/0408158].

 \bibitem{Cao:2004ap}
  Q.~H.~Cao, R.~Schwienhorst, and C.~P.~Yuan,
  Phys.\ Rev.\ D { 71}, 054023 (2005)
  [hep-ph/0409040].

\bibitem{Cao:2005pq}
  Q.~-H.~Cao, R.~Schwienhorst, J.~A.~Benitez, R.~Brock, C.~-P.~Yuan,
  Phys.\ Rev.\  {\bf D72}, 094027 (2005).
  [hep-ph/0504230].

\bibitem{Campbell:2005bb}
  J.~M.~Campbell and F.~Tramontano,
  Nucl.\ Phys.\  B {\bf 726}, 109 (2005)
  [arXiv:hep-ph/0506289].
  

\bibitem{Kidonakis:2006bu}
  N.~Kidonakis,
  Phys.\ Rev.\  D {\bf 74}, 114012 (2006)
  [arXiv:hep-ph/0609287].

 \bibitem{Zhang:2006cx}
  J.~J.~Zhang, C.~S.~Li, Z.~Li, L.~L.~Yang,
  Phys.\ Rev.\  {\bf D75}, 014020 (2007).
  [hep-ph/0610087].

\bibitem{Kidonakis:2007ej}
  N.~Kidonakis,
  Phys.\ Rev.\  D {\bf 75}, 071501 (2007)
  [arXiv:hep-ph/0701080].
  
\bibitem{Kidonakis:2008if} 
  N.~Kidonakis,
  Acta Phys.\ Polon.\ B {\bf 39}, 1593 (2008)
  [arXiv:0802.3381 [hep-ph]].

 \bibitem{Mirabella:2008gj}
  E.~Mirabella,
  Nuovo Cim.\  {\bf B123}, 1111-1117 (2008).
  [arXiv:0811.2051 [hep-ph]].
  
\bibitem{Kidonakis:2009sv} 
  N.~Kidonakis,
  Nucl.\ Phys.\ A {\bf 827}, 448C (2009)
  [arXiv:0901.2155 [hep-ph]].

\bibitem{Campbell:2009ss} 
  J.~M.~Campbell, R.~Frederix, F.~Maltoni and F.~Tramontano,
  Phys.\ Rev.\ Lett.\  {\bf 102}, 182003 (2009)
  [arXiv:0903.0005 [hep-ph]].
  
\bibitem{Kidonakis:2009ev} 
  N.~Kidonakis,
  Phys.\ Rev.\ Lett.\  {\bf 102}, 232003 (2009)
  [arXiv:0903.2561 [hep-ph]].

\bibitem{Campbell:2009gj} 
  J.~M.~Campbell, R.~Frederix, F.~Maltoni and F.~Tramontano,
  JHEP {\bf 0910}, 042 (2009)
  [arXiv:0907.3933 [hep-ph]].

\bibitem{Kidonakis:2009mx} 
  N.~Kidonakis,
  arXiv:0909.0037 [hep-ph].

\bibitem{Heim:2009ku}
  S.~Heim, Q.~-H.~Cao, R.~Schwienhorst, C.~-P.~Yuan,
  Phys.\ Rev.\  {\bf D81}, 034005 (2010).
  [arXiv:0911.0620 [hep-ph]].

 \bibitem{Ali:2009sm}
  A.~Ali, E.~A.~Kuraev, Y.~M.~Bystritskiy,
  Eur.\ Phys.\ J.\  {\bf C67}, 377-395 (2010).
  [arXiv:0911.3027 [hep-ph]].

 \bibitem{Gao:2009rf}
  J.~Gao, C.~S.~Li, J.~J.~Zhang, H.~X.~Zhu,
  Phys.\ Rev.\  {\bf D80}, 114017 (2009).
  [arXiv:0910.4349 [hep-ph]].

\bibitem{Kidonakis:2010tc}
  N.~Kidonakis,
  Phys.\ Rev.\  D {\bf 81}, 054028 (2010)
  [arXiv:1001.5034 [hep-ph]].

\bibitem{Kidonakis:2010ux} 
  N.~Kidonakis,
  Phys.\ Rev.\ D {\bf 82}, 054018 (2010)
  [arXiv:1005.4451 [hep-ph]].

 \bibitem{Bardin:2010mz}
  D.~Bardin, S.~Bondarenko, L.~Kalinovskaya, V.~Kolesnikov, W.~von Schlippe,
  Eur.\ Phys.\ J.\  {\bf C71}, 1533 (2011).
  [arXiv:1008.1859 [hep-ph]].

 \bibitem{Macorini:2010bp}
  G.~Macorini, S.~Moretti, L.~Panizzi,
  Phys.\ Rev.\  {\bf D82}, 054016 (2010).
  [arXiv:1006.1501 [hep-ph]].

\bibitem{Zhu:2010mr} 
  H.~X.~Zhu, C.~S.~Li, J.~Wang and J.~J.~Zhang,
  JHEP {\bf 1102}, 099 (2011)
  [arXiv:1006.0681 [hep-ph]].

\bibitem{Wang:2010ue} 
  J.~Wang, C.~S.~Li, H.~X.~Zhu and J.~J.~Zhang,
  arXiv:1010.4509 [hep-ph].
  
 \bibitem{Schwienhorst:2010je} 
  R.~Schwienhorst, C.~-P.~Yuan, C.~Mueller and Q.~-H.~Cao,
  Phys.\ Rev.\ D {\bf 83}, 034019 (2011)
  [arXiv:1012.5132 [hep-ph]].
   
\bibitem{Falgari:2010sf} 
  P.~Falgari, P.~Mellor and A.~Signer,
  Phys.\ Rev.\ D {\bf 82}, 054028 (2010)
  [arXiv:1007.0893 [hep-ph]].
  
\bibitem{Falgari:2011qa} 
  P.~Falgari, F.~Giannuzzi, P.~Mellor and A.~Signer,
  Phys.\ Rev.\ D {\bf 83}, 094013 (2011)
  [arXiv:1102.5267 [hep-ph]].
  
\bibitem{Kidonakis:2011wy}
  N.~Kidonakis,
  Phys.\ Rev.\  D {\bf 83}, 091503 (2011)
  [arXiv:1103.2792 [hep-ph]].

\bibitem{Kidonakis:2012db} 
  N.~Kidonakis,
  arXiv:1205.3453 [hep-ph].

  
\bibitem{Carlson:1993dt}
  D.~O.~Carlson and C.~P.~Yuan,
  Phys.\ Lett.\ B { 306}, 386 (1993).

\bibitem{Slabospitsky:2002ag}
  S.~R.~Slabospitsky and L.~Sonnenschein,
  Comput.\ Phys.\ Commun.\  { 148}, 87 (2002) 
  [hep-ph/0201292].


\bibitem{Herquet:2008zz}
  M.~Herquet, F.~Maltoni,
  Nucl.\ Phys.\ Proc.\ Suppl.\  {\bf 179-180}, 211-217 (2008).
  
\bibitem{Boos:2004kh}
  E.~Boos {\it et al.}, 
  Nucl.\ Instrum.\ Methods\ A { 534}, 250 (2004) 
  [hep-ph/0403113] 

\bibitem{Sjostrand:2006za}
  T.~Sjostrand, S.~Mrenna, P.~Z.~Skands,
  JHEP {\bf 0605}, 026 (2006).
  [hep-ph/0603175].  

\bibitem{Kersevan:2004yg}
  B.~P.~Kersevan, E.~Richter-Was,
  [hep-ph/0405247].

\bibitem{Frixione:2005vw}
  S.~Frixione, E.~Laenen, P.~Motylinski, B.~R.~Webber,
  JHEP {\bf 0603}, 092 (2006).
  [hep-ph/0512250].
  
\bibitem{Alioli:2009je}
  S.~Alioli, P.~Nason, C.~Oleari, E.~Re,
  JHEP {\bf 0909}, 111 (2009).
  [arXiv:0907.4076 [hep-ph]].  


\bibitem{Kane:1991bg}
  G.~L.~Kane, G.~A.~Ladinsky and C.~P.~Yuan,
  Phys.\ Rev.\  D {\bf 45}, 124 (1992).

\bibitem{Boos:1999dd}
  E.~Boos, L.~Dudko, and T.~Ohl,
  Eur.\ Phys.\ J.\ C { 11}, 473 (1999)
  [hep-ph/9903215].

\bibitem{AguilarSaavedra:2008zc}
  J.~A.~Aguilar-Saavedra,
  Nucl.\ Phys.\  B {\bf 812} (2009) 181
  [arXiv:0811.3842 [hep-ph]].

\bibitem{Zhang:2010dr}
  C.~Zhang and S.~Willenbrock,
  Phys.\ Rev.\  D {\bf 83}, 034006 (2011)
  [arXiv:1008.3869 [hep-ph]].

\bibitem{Tait:1999cf} 
  T.~M.~P.~Tait,
  Phys.\ Rev.\ D {\bf 61}, 034001 (2000)
  [hep-ph/9909352].
 
\bibitem{Belyaev:2000me} 
  A.~Belyaev and E.~Boos,
  Phys.\ Rev.\ D {\bf 63}, 034012 (2001)
  [hep-ph/0003260].

\bibitem{Boos:2006af}
  E.~E.~Boos, V.~E.~Bunichev, L.~V.~Dudko, V.~I.~Savrin, A.~V.~Sherstnev,
  Phys.\ Atom.\ Nucl.\  {\bf 69}, 1317-1329 (2006).
  
\bibitem{CERN-CMS-NOTE-2000-065}
  E.~Boos, L.~Dudko, V.~Savrin, CERN-CMS-NOTE-2000-065
  
\bibitem{Abazov:2001ns}
  V.~M.~Abazov {\it et al.} [ D0 Collaboration ],
  Phys.\ Lett.\  {\bf B517}, 282-294 (2001).
  [hep-ex/0106059].

  B.~Abbott {\it et al.} [ D0 Collaboration ],
  Phys.\ Rev.\  {\bf D63}, 031101 (2000).
  [hep-ex/0008024].

  D.~Acosta {\it et al.} [ CDF Collaboration ],
  Phys.\ Rev.\  {\bf D65}, 091102 (2002).
  [hep-ex/0110067].

  D.~Acosta {\it et al.} [ CDF Collaboration ],
  Phys.\ Rev.\  {\bf D69}, 052003 (2004).

\bibitem{Abazov:2006gd}
  V.~M.~Abazov {\it et al.} [ D0 Collaboration ],
  Phys.\ Rev.\ Lett.\  {\bf 98}, 181802 (2007).
  [hep-ex/0612052].

  V.~M.~Abazov {\it et al.} [ D0 Collaboration ],
  Phys.\ Rev.\  {\bf D78}, 012005 (2008).
  [arXiv:0803.0739 [hep-ex]].
 
\bibitem{Aaltonen:2008sy}
  T.~Aaltonen {\it et al.} [ CDF Collaboration ],
  Phys.\ Rev.\ Lett.\  {\bf 101}, 252001 (2008).
  [arXiv:0809.2581 [hep-ex]].

\bibitem{Abazov:2011pt}
  V.~M.~Abazov {\it et al.} [ D0 Collaboration ],
  [arXiv:1108.3091 [hep-ex]].
  
\bibitem{Aaltonen:2010jr}
  T.~Aaltonen {\it et al.} [ CDF Collaboration ],
  Phys.\ Rev.\  {\bf D82}, 112005 (2010).
  [arXiv:1004.1181 [hep-ex]].



\bibitem{Chatrchyan:2011vp}
  S.~Chatrchyan {\it et al.} [ CMS Collaboration ],
  Phys.\  Rev.\  Lett.\  {\bf 107}, 091802 (2011).
  [arXiv:1106.3052 [hep-ex]].

\bibitem{CMS-t-channel}
The CMS Collaboration, CMS-PAS-TOP-11-021

\bibitem{CMS-tw} 
  S.~Chatrchyan {\it et al.}  [ CMS Collaboration],
  arXiv:1209.3489 [hep-ex], CMS-PAS-TOP-11-022.



\bibitem{ATLAS-t}
The ATLAS collaboration, ATLAS-CONF-2011-101


\bibitem{:2012dj} 
  G.~Aad {\it et al.}  [ATLAS Collaboration],
  Phys.\ Lett.\ B {\bf 716}, 142 (2012)
  [arXiv:1205.5764 [hep-ex]].
  
\bibitem{ATLAS-s}
The ATLAS collaboration, ATLAS-CONF-2011-118
``Search for s-Channel Single Top-Quark Production in Collisions at $\sqrt{s} = 7$ TeV''


\bibitem{Boos:2001sj}
  E.~Boos, M.~Dubinin, A.~Pukhov, M.~Sachwitz, H.~J.~Schreiber,
  Eur.\ Phys.\ J.\  {\bf C21}, 81-91 (2001).

\bibitem{Bardin}
D.~Bardin, R.~Kleiss et. al., in: {\it Physics at LEP2}, ed.by 
G.Altarelli, T.Sjoestrand, F.Zwirner, CERN report 96-01, 1996, vol. II
(hep-ph/9709270)

\bibitem{sensitivity}
E.~Boos, Y.~Kurihara, M.~Sachwitz, H.J.~Schreiber, S.~Shichanin,
Y.~Shimizu, Z.Phys {\bf C70} (1996) 255

\bibitem{boos-ohl}
E.~Boos, T.~Ohl,
Phys.Rev.Lett. {\bf 83} (1999) 480
(hep-ph/9903357)

\bibitem{Kuhn:2003pn}
J.~H.~Kuhn, C.~Sturm and P.~Uwer,
arXiv:hep-ph/0303233.

 \bibitem{Penunuri:2011hp}
  F.~Penunuri, F.~Larios, A.~O.~Bouzas,
  Phys.\ Rev.\  {\bf D83}, 077501 (2011).
  [arXiv:1102.1417 [hep-ph]].


\bibitem{Weiglein:2004hn}
  G.~Weiglein {\it et al.}  [LHC/LC Study Group],
  Phys.\ Rept.\  {\bf 426}, 47 (2006)
  [arXiv:hep-ph/0410364].


\bibitem{Buchmuller:1985jz} 
  W.~Buchmuller and D.~Wyler,
  Nucl.\ Phys.\ B {\bf 268}, 621 (1986).

\bibitem{Zhang:2012cd} 
  C.~Zhang, N.~Greiner, and S.~Willenbrock,
  arXiv:1201.6670 [hep-ph];
  C.~Zhang and S.~Willenbrock,
  Phys.\ Rev.\ D {\bf 83}, 034006 (2011).
 

\bibitem{Tsuno:2005qb} 
  S.~Tsuno, I.~Nakano, Y.~Sumino and R.~Tanaka,
  Phys.\ Rev.\ D {\bf 73}, 054011 (2006)
  [hep-ex/0512037].
 
\bibitem{AguilarSaavedra:2008gt} 
  J.~A.~Aguilar-Saavedra,
  Nucl.\ Phys.\ B {\bf 804}, 160 (2008)
  [arXiv:0803.3810 [hep-ph]].

\bibitem{Bernreuther:2008us} 
  W.~Bernreuther, P.~Gonzalez and M.~Wiebusch,
  Eur.\ Phys.\ J.\ C {\bf 60}, 197 (2009)
  [arXiv:0812.1643 [hep-ph]].
 
\bibitem{Boos:1997rd} 
  E.~Boos, A.~Pukhov, M.~Sachwitz and H.~J.~Schreiber,
  Phys.\ Lett.\ B {\bf 404}, 119 (1997)
  [hep-ph/9704259].

\bibitem{Chen:2005vr} 
  C.~-R.~Chen, F.~Larios and C.~-P.~Yuan,
  Phys.\ Lett.\ B {\bf 631}, 126 (2005)
  [AIP Conf.\ Proc.\  {\bf 792}, 591 (2005)]
  [hep-ph/0503040].
  
\bibitem{:2012iwa} 
  V.~M.~Abazov {\it et al.}  [D0 Collaboration],
  Phys.\ Lett.\ B {\bf 713}, 165 (2012)
  [arXiv:1204.2332 [hep-ex]].

\bibitem{Drobnak:2011aa} 
  J.~Drobnak, S.~Fajfer and J.~F.~Kamenik,
  Nucl.\ Phys.\ B {\bf 855}, 82 (2012);

\bibitem{Han:1998tp} 
  T.~Han, M.~Hosch, K.~Whisnant, B.~-L.~Young and X.~Zhang,
  Phys.\ Rev.\ D {\bf 58}, 073008 (1998)
  [hep-ph/9806486].

\bibitem{Liu:2005dp} 
  J.~J.~Liu, C.~S.~Li, L.~L.~Yang and L.~G.~Jin,
  Phys.\ Rev.\ D {\bf 72}, 074018 (2005)
  [hep-ph/0508016].
  
\bibitem{Abazov:2010qk} 
  V.~M.~Abazov {\it et al.}  [D0 Collaboration],
  Phys.\ Lett.\ B {\bf 693}, 81 (2010)
  [arXiv:1006.3575 [hep-ex]].

\bibitem{Aaltonen:2008qr} 
  T.~Aaltonen {\it et al.}  [CDF Collaboration],
  Phys.\ Rev.\ Lett.\  {\bf 102}, 151801 (2009)
  [arXiv:0812.3400 [hep-ex]].
  
\bibitem{Aad:2012gd} 
  G.~Aad {\it et al.}  [ATLAS Collaboration],
  Phys.\ Lett.\ B {\bf 712}, 351 (2012)
  [arXiv:1203.0529 [hep-ex]].


\bibitem{Chivukula:1995gu}
R.~S.~Chivukula, E.~H.~Simmons and J.~Terning,
Phys.\ Rev.\ D {\bf 53}, 5258 (1996).

\bibitem{Kaplan:1983fs}
D.~B.~Kaplan and H.~Georgi,
Phys.\ Lett.\ B {\bf 136}, 183 (1984).

\bibitem{Georgi:1984af}
H.~Georgi and D.~B.~Kaplan,
Phys.\ Lett.\ B {\bf 145}, 216 (1984).

\bibitem{Arkani:2001nc}
N.~Arkani-Hamed, A.~G.~Cohen and H.~Georgi,
Phys.\ Lett.\ B {\bf 513}, 232 (2001).

\bibitem{Kaplan:2003uc}
D.~E.~Kaplan and M.~Schmaltz,
JHEP {\bf 0310}, 039 (2003).

\bibitem{Schmaltz:2005ky}
  M.~Schmaltz and D.~Tucker-Smith,
  Ann.\ Rev.\ Nucl.\ Part.\ Sci.\  {\bf 55}, 229 (2005).


\bibitem{composites}
B. Schrempp,
Proceedings of the 23rd International Conference on High Energy Physics,
Berkeley (World Scientific, Singapore 1987);  

U. Baur {\it et al.}, Phys. Rev. {\bf D35}, 297 (1987);

M. Kuroda {\it et al.}, Nucl. Phys. {\bf B261}, 432 (1985).


\bibitem{Batra:2003nj}
  P.~Batra, A.~Delgado, D.~E.~Kaplan and T.~M.~P.~Tait,
  JHEP {\bf 0402}, 043 (2004)


\bibitem{guts}
R.W. Robinett, Phys. Rev. {\bf D26}, 2388 (1982);

R.W. Robinett and J.L. Rosner, Phys. Rev. {\bf D26}, 2396 (1982);

P. Langacker, R. W. Robinett, and J.L. Rosner, Phys. Rev. {\bf
D30}, 1470 (1984).


\bibitem{Cvetic:1996mf}
  M.~Cvetic and P.~Langacker,
  Mod.\ Phys.\ Lett.\ A {\bf 11}, 1247 (1996).

\bibitem{Pati:2006nw}
J.~C.~Pati,
  arXiv:hep-ph/0606089.

\bibitem{superstrings}
M. Green and J. Schwarz, Phys. Lett. {\bf 149B}, 117(1984); 

D. Gross {\it et al.}, Phys. Rev. Lett. {\bf 54}, 502 (1985);

E. Witten, Phys. Lett. {\bf 155B}, 1551 (1985); 

P. Candelas {\it et al.}, Nucl. Phys. {\bf B258}, 46 (1985);

M. Dine {\it et al.}, Nucl. Phys. {\bf B259}, 549 (1985).

\bibitem{Pati:1974yy}
J.~C.~Pati and A.~Salam,
Phys.\ Rev.\ D {\bf 10}, 275 (1974).

\bibitem{Mohapatra:1974hk}
R.~N.~Mohapatra and J.~C.~Pati,
Phys.\ Rev.\ D {\bf 11}, 566 (1975).

\bibitem{Mohapatra:1974gc}
R.~N.~Mohapatra and J.~C.~Pati,
Phys.\ Rev.\ D {\bf 11}, 2558 (1975).

\bibitem{Senjanovic:1975rk}
G.~Senjanovic and R.~N.~Mohapatra,
Phys.\ Rev.\ D {\bf 12}, 1502 (1975).

\bibitem{Mimura:2002te}
Y.~Mimura and S.~Nandi,
Phys.\ Lett.\ B {\bf 538}, 406 (2002).

\bibitem{Cvetic:1983su}
M.~Cvetic and J.~C.~Pati,
Phys.\ Lett.\ B {\bf 135}, 57 (1984).

\bibitem{Langacker:1989xa}
P.~Langacker and S.~Uma Sankar,
Phys.\ Rev.\ D {\bf 40}, 1569 (1989).


\bibitem{Datta:2000gm}
A.~Datta, P.~J.~O'Donnell, Z.~H.~Lin, X.~Zhang and T.~Huang,
Phys.\ Lett.\ B {\bf 483}, 203 (2000).

\bibitem{extradimensions}
R.~Sundrum, 
arXiv:hep-th/0508134;

C.~Csaki, Jay~Hubisz, Patrick~Meade, 
arXiv:hep-ph/0510275;

G.~Kribs, 
arXiv:hep-ph/0605325.

\bibitem{Sullivan:2002jt} 
  Z.~Sullivan,
  Phys.\ Rev.\ D {\bf 66}, 075011 (2002)
  [hep-ph/0207290].
 
\bibitem{Boos:2011ib} 
  E.~E.~Boos, I.~P.~Volobuev, M.~A.~Perfilov and M.~N.~Smolyakov,
  Theor.\ Math.\ Phys.\  {\bf 170}, 90 (2012)
  [arXiv:1106.2400 [hep-ph]].
 
\bibitem{Boos:2006xe} 
  E.~Boos, V.~Bunichev, L.~Dudko and M.~Perfilov,
  Phys.\ Lett.\ B {\bf 655}, 245 (2007)
  [hep-ph/0610080].

\bibitem{Abazov:2011xs} 
  V.~M.~Abazov {\it et al.}  [D0 Collaboration],
  Phys.\ Lett.\ B {\bf 699}, 145 (2011)
  [arXiv:1101.0806 [hep-ex]].

\bibitem{Aaltonen:2009qu} 
  T.~Aaltonen {\it et al.}  [CDF Collaboration],
  Phys.\ Rev.\ Lett.\  {\bf 103}, 041801 (2009)
  [arXiv:0902.3276 [hep-ex]].
  
\bibitem{CMS-PAS-EXO-12-001} 
  S.~Chatrchyan {\it et al.}  [ CMS Collaboration],
  CMS-PAS-EXO-12-001
  
\bibitem{Aad:2012ej} 
  G.~Aad {\it et al.}  [ATLAS Collaboration],
  Phys.\ Rev.\ Lett.\  {\bf 109}, 081801 (2012)
  [arXiv:1205.1016 [hep-ex]].


\bibitem{Abazov:2008rn} 
  V.~M.~Abazov {\it et al.}  [D0 Collaboration],
  Phys.\ Rev.\ Lett.\  {\bf 102}, 191802 (2009)
  [arXiv:0807.0859 [hep-ex]].

  
\end{thebibliography}
\end{document}